\documentclass[aps,prd,showpacs,showkeys,preprint,amsmath,amssymb]{revtex4-1}
\usepackage[english]{babel} %,portuguese
\usepackage{epsfig,float,subfigure,amssymb,amsmath,amsthm} %,t1enc
\makeatletter

\newcommand{\beq}{\begin{equation}}
\newcommand{\eeq}{\end{equation}}
\newcommand{\bea}{\begin{eqnarray}}
\newcommand{\eea}{\end{eqnarray}}

\usepackage{amssymb,amsmath}

\def\cal{\mathcal}
\DeclareMathOperator{\sech}{sech}

\begin{document}
\title{Confluent Heun functions in gauge theories on thick braneworlds }
\author{M. S. Cunha$^\dag$ and H. R. Christiansen$^\dag {}^*$}
\affiliation{$^\dag$ Grupo de F\'{\i}sica Te\'orica, State University of Ceara (UECE), Av. Paranjana 1700, 60740-903 Fortaleza - CE, Brazil\\
$^*$ Universidade Estadual Vale do Acara\'{u}, Av. da Universidade 850, 62040-370 Sobral - CE, Brazil }

\begin{abstract}
We investigate the propagation modes of gauge fields in an infinite %noncompact
Randall-Sundrum %type, like
scenario. In this model a sine-Gordon soliton represents our \textit{thick} four-dimensional braneworld while an exponentially coupled scalar acts for the dilaton field. For the gauge-field motion we find a differential equation which can be transformed into a confluent Heun equation. By means of another change of variables we obtain a related Schrodinger equation with a family of symmetric rational  $(\gamma-\omega z^2)/(1-z^2)^2$ potential functions.  We discuss both results and present the infinite spectrum of analytical solutions for the gauge field. Finally, we assess the existence and the relative weights of Kaluza-Klein modes in the present setup.

\end{abstract}

11.10.Kk., 04.50.-h. \hfill \texttt{}

%Keywords: Large Extra Dimensions, Field Theories in Higher Dimensions

\maketitle
\section{Introduction}

One of the main purposes of superstring theory is the inclusion of
all the relevant fields of Nature together in one single Lagrangian.
Field theoretic scenarios inspired in such a theory put in contact
gauge and matter fields with metric degrees of freedom, altogether defined
in some extra-dimensional space \cite{polchinski}.
Extra-dimensions, combined with the influence of the gravitational field,
modify nontrivially all sectors so that gauge forces
must be proven to remain the same in the usual four-dimensional (4D) subspace or predict new physics
in some consistent way.
Localization in the gauge sector is expected to hold and
the effective 4D electromagnetic force must be mediated by massless photons as usual.
On the other hand, higher-dimensional spaces help solving fundamental
problems such as the hierarchy gap between the Planck and gauge-coupling scales 
in the Standard model \cite{RS}.

Also inspired in String theory is the use of {branes} to represent
our Universe. In String theory, gauge modes are deposited on D-branes
from open strings ending on them, so we expect gauge fields
in field-theoretic models to have finite localized modes on  {stringy}
topological defects of lower dimensionality. Actually, to obtain finite
eigenstates in 4D it has been shown that to fulfill this task  we need not
just gravity but also a dilaton \cite{youm1, keha-tamva}, a field already
predicted in String theory.

In the present paper, we describe gauge fields
in a warped five-dimensional bulk with a dilaton and a
brane defect that mimics the ordinary world. Both brane and dilaton configurations
are geometrically consistent solutions of a two scalar world action in 
a curved 5D space-time where the field potential is of sine-Gordon type.

We show a relationship between the fundamental parameters of the 5D theory
which is crucial to determine the dynamics of the fields both in the bulk
and  ordinary space. Indeed, for different choices of a parameter defined by
the quotient of some power of the sine-Gordon frequency-amplitude and the 5D Planck mass, the
equations of motion of the gauge field can be completely different.
Notably, in Ref. \cite{CHRISTIANSEN} we have been able to find the whole spectrum of a theory
involving both Maxwell and Kalb-Ramond fields for a particular value of this parameter.
As we will see, there exists a minimal value for the dilaton coupling constant above which
the finiteness of the action is assured and it is directly related to the localization of gauge fields.

In what follows we analytically obtain the propagation modes (massless and massive) of a
gauge theory in a background of the sine-Gordon type that results in new equations of motion.
We show that the dynamics of the quantum mechanical system associated with the
problem is given by a simple (rational) potential function and that
the solutions to the Schrodinger equation are of the Mathieu type (with a power-law factor).
In a more general case we obtain the exact spectrum
given by the set of confluent Heun functions and show that Kaluza-Klein
states are strongly suppressed in ordinary space.

The paper is organized as follows. In the next Section, we present the geometrical
background. In Section \ref{sect action} we introduce the action for the 5D gauge field
coupled to warped gravity and a dilaton background and
derive the 5D equations of motion.
In Section \ref{quantum analog} we obtain the quantum analog problem
showing explicitly the quantum-mechanical Schrodinger potential. The eigenvalue
spectrum is computed and graphically shown. Next, in Sections \ref{heun} and \ref{remarks}
we discuss the general problem and draw our conclusions.
Other recent results about thick braneworlds can be found in e.g. \cite{recent}.

\section{Geometrical background\label{sect background}}

Our framework is a five-dimensional space-time embedding a
four-dimensional membrane also called thick brane.
The (space-like) extra-dimension is assumed infinite and the brane
will be dynamically obtained as a solution to the Einstein equations
for gravity coupled to a pair of scalar fields. One of these scalars represents
a domain wall defect (the thick brane) while the other is the dilaton.
The dilaton, together with the warping of the fifth dimension,
happens to be crucial in the gauge theory that will be developed and makes
more clear the stringy origin of the theory.
Since gauge field theory is conformal \cite{dvali shifman} all the information coming
from the warping of the 4D metric is automatically lost. As a consequence the photon
is non-normalizable in the four-dimensional space unless the gauge coupling is dynamically modified. Indeed, the exponential coupling of both the dilaton and the 5D warping to
the gauge field conveniently modifies the scaling properties and the zero-mode
becomes localized \cite{keha-tamva}.

The five-dimensional world action which determines the background is
\begin{equation}
S_B=\int d^{4}x\ dy
\sqrt{-\det G_{MN}}\ [2M^{3}R-\frac{1}{2}(\partial\Phi)^{2}-\frac{1}{2}%
(\partial\Pi)^{2}-\mathcal{V}(\Phi,\Pi)],
\label{action bounce}
\end{equation}
where $M$ is the Planck mass in 5D, and $R$ is the Ricci scalar.
The solution for $\Phi$ represents the world membrane and
the corresponding field solution for $\Pi$ will be the dilaton configuration
consistent with the metric and the kink.
As usual we adopt Latin capitals on the bulk and Greek lower case letters on 4D.

We next adopt the following ansatz for the metric
\begin{equation}
ds^{2}=e^{2\Lambda(y)}\eta_{\mu\nu}dx^{\mu}dx^{\nu}+e^{2\Sigma(y)}dy^{2},
\label{warpedmetric}
\end{equation}
where $\Lambda$ and $\Sigma$ depend just on
the fifth coordinate, $y$,  and diag$(\eta)=(-1, 1, 1, 1)$. The
equations of motion for eq.(\ref{action bounce}) are
\bea &&\frac{1}{2}(\Phi^{\prime})^{2}+\frac{1}{2}(\Pi^{\prime})^{2}-
e^{2\Sigma(y)}\mathcal{V}(\Phi,\Pi)=24M^{3}(\Lambda^{\prime})^{2},\label{motion1}\\
&&\frac{1}{2}(\Phi^{\prime})^{2}+\frac{1}{2}(\Pi^{\prime})^{2}+e^{2\Sigma(y)}\mathcal{V}(\Phi,\Pi)=
-12M^{3}\Lambda^{\prime\prime}-24M^{3}(\Lambda^{\prime})^{2}+12M^{3}\Lambda^{\prime}\Sigma^{\prime},
\nonumber\eea and  \bea
\Phi^{\prime\prime}+(4\Lambda^{\prime}-\Sigma^{\prime})\Phi^{\prime}=
e^{2\Sigma}\ \frac{\partial\mathcal{V}}
{\partial\Phi},\nonumber\\
\Pi^{\prime\prime}+(4\Lambda^{\prime}-\Sigma^{\prime})\Pi^{\prime}=
e^{2\Sigma}\ \frac{\partial\mathcal{V}}{\partial\Pi}.
\label{motion2} \eea where the prime means derivative with respect to $y$.

By means of a supergravity motivated functional $\mathcal{W}(\Phi)$ defined by
\beq\Phi^{\prime}=\frac{d\mathcal{W}}{d\Phi}\label{W}\eeq
%so-called superpotential
\cite{supergravity} the system of differential equations can be more easily handled.
This method is also applicable
to non-supersymmetric domain walls \cite{superpot back, superpotential} as the present one.

First, we consider
the action in the absence of gravity (and no dilaton) in order to obtain
an expression for $\Phi$. Then, we put this solution into the equations
of motion (\ref{motion1}) and (\ref{motion2}).

The standard sine-Gordon Lagrangian reads
\beq L_{SG}=-\frac 12\partial^2\Phi- V(\Phi) \eeq
with $$V(\Phi)=\frac 1 {b^2}\ (1-\cos (b\,\Phi)).$$
The free parameter $b$ signals bulk-symmetries $\delta\Phi\rightarrow 2n \pi/b\ (n  \in Z)$ among the
vacua of this theory. Solutions interpolating vacua are possible and,
assuming they depend only on $y$, one-solitons read
\beq
\Phi(y)=\frac 4 b \ \arctan\ e^y.\label{newbounce}\eeq
These functions kink on our 4D-world slice, namely at $y\sim0$.

In a gravitational background of the form  (\ref{warpedmetric}), now including also the dilaton,
the equations of motion (\ref{motion1}) and (\ref{motion2}) are still
compatible with  solutions (\ref{newbounce}) provided we find the appropriate
potential functional $\mathcal{V}$ for the general action (\ref{action bounce}), viz.
\begin{equation}
\mathcal{V}(\Phi,\Pi)=\exp{({\Pi}/{\sqrt{12M^{3}}})}
\left(\frac{1}{2}(\frac{d\mathcal{W}}{d\Phi})^{2}-\frac{5}{32M^{3}}\mathcal{W}(\Phi)^{2}\right).
\label{potential}
\end{equation}
Taking into account eq.(\ref{W}), the superpotential functional results
\begin{equation}
\mathcal{W}(\Phi)=-\frac 4{b^2}\cos(\frac {b}2\,\Phi)
\end{equation}
and then
\begin{equation}
\mathcal{{V}}(\Phi, \Pi)=-e^{({\Pi}/{\sqrt{12M^{3}}})}
\left(\frac 4{b^2}\sin^2\frac b 2\,\Phi+\frac{5}{2M^3 b^4}\cos^{2}\frac b 2\,\Phi\right).
\label{newpotential}
\end{equation}
If we now conveniently write the Hamiltonian \`a la Bogomol'nyi,
we can detect the following relations among the warping functions,
the dilaton and the superpotential
\beq
\Pi=-\sqrt{3M^3}\Lambda, \ \ \Sigma=\Lambda/4, \ \ \Lambda^\prime=-
\mathcal{W}/12M^3. \label{newsol}\eeq
Finally, totally solving the equations of motion,
the dilaton field is given by
\beq
\Pi(y)= \frac 1 {\sqrt{3M^3} b^2} \ln\cosh y\label{newsolution},\eeq
and
\beq
\Lambda=-\frac 1 {{3M^3} b^2} \ln\cosh y, \ \ \Sigma=-\frac 1 {{12M^3} b^2} \ln\cosh y.
 \label{warp}\eeq

\begin{figure}[h]
 \centering %here we have set $M=1$
\includegraphics[width=7.5cm,height=5.5cm]{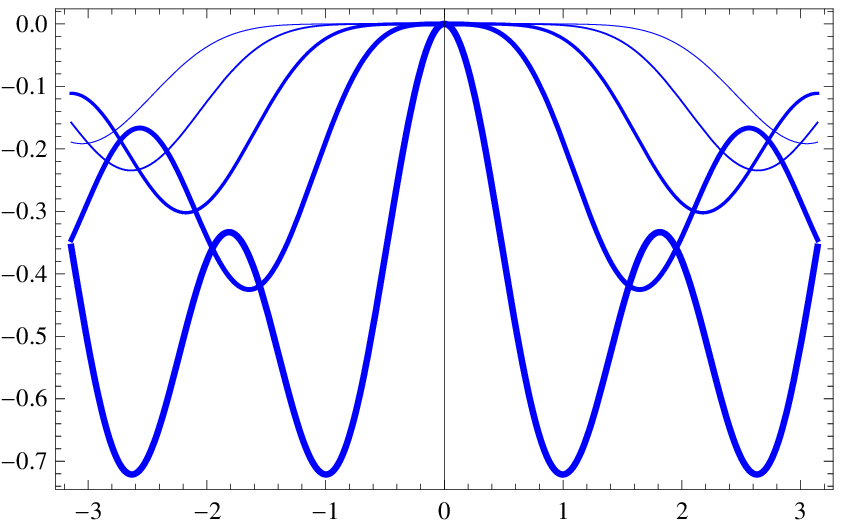}
%\caption{\label{potphipar}}
\includegraphics[width=7.5cm,height=5.5cm]{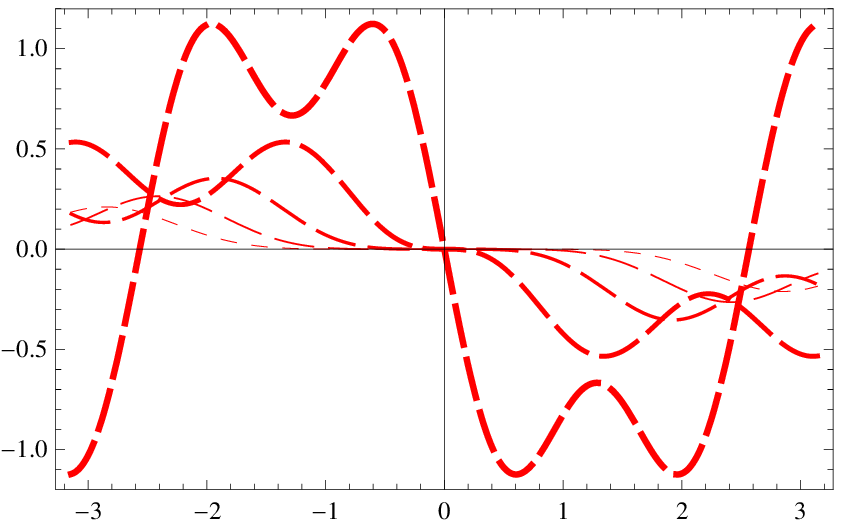}
\caption {\label{potphi} Family of background potential functionals $\mathcal{{V}}(\Phi)$
(eq.\ref{newpotential phi}) for
different values of $a\equiv1/6M^3b^2$: even $a=2, 4, 6, 8, 10$ (solid line), odd $a=1, 3, 5, 7, 9$ (dashed line).
For clarity we adopted thinner lines for bigger values of $a$.}
\end{figure}

The relation between $\Pi$ and $\Phi$ allows also writing $\mathcal{{V}}$ as
\begin{equation}
\mathcal{{V}}(\Phi)=-\frac 4 {b^2}\ (\sin \frac b 2 \Phi)^{1/6M^3{b^2}}\
\left(1+(\frac{5}{8M^3 b^2}-1)\cos^{2}\frac b 2\,\Phi\right),
\label{newpotential phi}
\end{equation}
which fully shows its dependence on $b$ and $M$ (see Fig. \ref{potphi}).
\bigskip
As it happens with dilaton configurations related to D-brane solutions,  functions
such as (\ref{newbounce}) and (\ref{newsolution}) are singular when $|y|\rightarrow\infty$.
However, since the metric vanishes exponentially and both dilaton and warp factors
operate under an exponential coupling, the model is kept free of divergences.

The warping functions amount to a
shift in the effective four-dimensional Planck scale,
which remains finite with the following definition
\beq M_P^2\equiv M^3\int_{-\infty}^\infty dy \ e^{4\Lambda(y)+\Sigma(y)}.\eeq

Using the consistency relation (\ref{newsol}) just found, the action reads
\beq S_B\sim\int dy\ e^{4\Lambda(y)+\frac 1 4\Lambda(y)+\frac{\lambda}2\sqrt{3M^3}\Lambda(y)}\ S_{(4)}
\eeq
where $S_{(4)}$ is the remaining of the action integrated in 4D. According to the solution
$
\Lambda(y) = 2a\ln \rm sech y $ (c.f. eq.(\ref{newsol}) - eq.(\ref{warp}))
the 5D factor results finite
provided $c\equiv(17+2\lambda\sqrt{3M^3})/4>0$, namely $\lambda> -\frac{17}{2\sqrt{3M^3}}=\lambda_0$.

Studying the fluctuations of the metric about the above background configuration,
it is possible to see that this model supports a massless zero-mode of the gravitational
field localized on the membrane even in the presence of the dilaton. In order to prove the
stability of the background solution,
we would have to show that there are no negative mass solutions to the equations of motion
of a perturbation $h_{\mu\nu}$ of the metric.
Actually, a gravitational Kaluza-Klein spectrum appears, starting from zero and presenting no gap.
This can be easily seen after an appropriate change of variables
and decomposition of the gravitational field, and a subsequent supersymmetric type expression
of the Schrodinger type operator resulting from the equation of motion
(see \cite{keha-tamva, gremm} for details).
The issue of the coupling of these massive modes to the brane has been analyzed in detail
in \cite{csaki al}.

\section{Gauge field action in a warped space with dilaton \label{sect action}}

Let us consider the following 5D action where a five-dimensional electromagnetic field $A_{N}$
is coupled to the dilaton \cite{mayr} in a warped space-time

\begin{equation}
S_g = \int dy\,d^4x\,\sqrt{-\det G_{AB}}\ e^{-\frac\lambda
2 \Pi }\ \left\{-\dfrac14\,F_{MN}F^{MN} \right\}
\label{action gauge}
\end{equation}
where $F_{MN}=\partial_{[M}A_{N]}$.

Assuming that the gauge field energy density should not
strongly modify the geometrical background, we can study the
behavior of the propagating modes in the background of the topological
configuration  studied in the last Section.
In general, most of the attempts to stabilize 5D brane worlds by means of a scalar field in
the bulk do not take into account the back-reaction of the scalar field on the
background metric \cite{keha-tamva, scalar RS, gremm, csaki al} and those in order to compute the
scalar back-reaction on the metric
were unsuccessful except in a few special cases \cite{superpot back, scalar backreaction}.

The factor $\exp(\Sigma-\lambda\Pi/2)$, present in the integrand, will lead to a change
in the integration measure which is crucial to conserve the effect of the warping function
on the 4D gauge field. As a consequence, zero-modes (namely photons)
are normalizable in the 4D effective theory. To see it in detail, we need to solve the 5D
equations of motion for $A_M$
\beq
\frac 1{\sqrt{-G}}\,\partial_M (G^{MR}G^{NP}F_{RP} \sqrt{-G} e^{-\frac \lambda 2\Pi(y)})=0
\label{more motion}
\eeq
where diag $G_{MN}=(e^{2\Lambda}\eta_{\mu\nu}, e^{2\Sigma})$.
For this, we adopt the following gauge choice
$ A^{5}=0,\ \partial_{\mu}A^{\mu}=0$
and separate the fifth
from the ordinary coordinates as follows
\beq A^{\mu}(x, y)=a^{\mu}(x)u(y).\label{decomposition}\eeq
%with $\partial_5=\partial/\partial y, \partial_{\mu}=\eta_{\mu\nu}\partial_\mu$ and $a^{\mu}=\eta^{\mu\nu}a_\mu$. .

Now, from eq.(\ref{more motion}) we just get
\beq
  [\Box +\frac 1{u\ f} \partial_5(f\partial^5 u)]\ a^{\mu}=0.\label{motion5}
\eeq
Note that the warped metric and the dilaton deform the solutions of this
differential equation by means of the factor
$f(y)\equiv e^{4\Lambda+\Sigma-\lambda\Pi/2}$ multiplying $u(y)$ and $u'(y)$.
A full Kaluza-Klein spectrum results from the solution of the general case
\beq \partial_5(f\partial^5 u)=-m^2 f u, \eeq
where $m^2$ is an arbitrary constant representing the 4D squared boson mass
of the vector gauge field. It means that $a^{\mu}=a^{\mu}(0)e^{ipx}$
with $p^2=-m^2$.

By expanding eq.(\ref{motion5}), we obtain the most general $y$-dependent equation
of motion for the modified sine-Gordon potential
(\ref{newpotential phi}) derived from action (\ref{action gauge}) in a form which exhibits
its dependence on $a=1/6M^3b^2$,  and $c$
\beq
 u''(y)+ a (1-2c)\tanh y\ u'(y) + m^2\sech^{a}y\ u(y)=0\label{massive modes},\eeq
where $ y\in(-\infty, \infty)$ as already stated.
Looking back at the definition of the auxiliary constants we get the explicit
dependence of the solutions on the original parameters $b$, $\lambda$ and $M$.

\ {}
\ {}

Below, we will discuss the possible values of $m^2$ as resulting from an eingenvalue problem related to the equation of motion (\ref{massive modes}).
Indeed, there exists a Schrodinger like equation
equivalent to eq.(\ref{massive modes}) with a
potential function which concentrates all the richness implicit in the complicated
equation (\ref{massive modes}).
Note that the particular solution $u(y)=$constant represents
the $m^2 = 0$ \textit{photon} state of the 5D theory.
Since this solution satisfies eq.(\ref{massive modes})
for any value of $a$ and $c$, any member of the family of problems has a guaranteed
localized zero-mode. See \cite{CHRISTIANSEN} for details.

Localization of  gauge-field modes in the ordinary space
can be established by verifying that the corresponding
5D action is finite. From eq.(\ref{decomposition}) one has
$F^{\mu\nu}=f^{\mu\nu}u(y)$, where $f^{\mu\nu}=G^{\mu\alpha}G^{\nu\beta}f_{\alpha\beta}$,
so that for a gauge mode $A_M^{\rm sol.}$ the relevant part of eq.(\ref{action gauge}) reads
\beq
S_g [A_M^{\rm sol.}] = \int dy\ u^{2}(y)e^{4\Lambda(y)+\Sigma(y)-\lambda\Pi(y)/2}\int
d^{4}x\ \frac 1 4 f_{\mu\nu}f^{\mu\nu}.\label{finiteaction}
\eeq
Using the field solutions found in eq.(\ref{newbounce}) and the equations thereafter,
the fifth dimension factor 
will remain finite for each mode $u(y)$ growing below $e^{a c}$ at infinity. Thus,
any finite solution  is a physically acceptable Kaluza-Klein state (as we have seen above,
to have a finite 5D Planck mass and background action $S_B$
we already need $c > 0$, i.e. $\lambda> \lambda_0$).

It is known that by means of a transformation
\beq  u(y)=e^{-\alpha \Lambda/2}U(z),\ \ \ \frac{dz}{dy}= e^{-\beta \Lambda}\label{change variables} \eeq
we can turn eq.(\ref{massive modes}) into a Schrodinger-like equation in the variable $z$
(see e.g. \cite{RS, keha-tamva}). In general, the existence of an analog Schrodinger equation is useful to give us a feeling of the physical profile of the solutions of the original problem, as e.g. parity and eigenvalues.
With $\alpha=c-1/4$ and $\beta=-1/4$ we can eliminate the first derivative term in $U$ and have a pure mass term as usual. The resulting differential equation reads precisely
\beq \left[-\frac{d^2}{dz^2}+ \mathfrak{V}_a(z)\right]U(z)=m^2U(z),\nonumber\eeq
where $\mathfrak{V}_a(z)=e^{-\Lambda/2}(\frac{\alpha}2\Lambda''-\gamma\Lambda'^2)$ and
$\gamma=\frac 1 4\alpha(\frac 1 2-\alpha)$.
% derivatives are with respect to $y$ (we eventually replace y(z))
In a few cases the last expression can be inverted after exact integration
in order that an analytical expression for the analog non-relativistic potential comes about.
In Ref.\cite{CHRISTIANSEN} we have solved the $a=2$ case and found
\beq \mathfrak{V}_2(z)= -2\alpha \left[1-(2\alpha-1) \tan^2z\right].
\label{potential2}\nonumber \eeq In this paper, for $a=4$ we find
\beq \mathfrak{V}_4(z)= \frac{(\gamma-\omega z^2)}{(1-z^2)^2},
\nonumber\eeq  where $\gamma$ and $\omega$ are constants and we shall analyze it in what follows.

%%%%%%%%%%%%%%%%%%%%%%%%%%%%%%%%%%%%%%%%%%
\section{The quantum analog \label{quantum analog}}

In the present case we can turn eq.(\ref{massive modes}) into a Sturm-Liouville problem by means of
\beq
z=\tanh\!y,
\eeq
\beq
u(y)=(\cosh \!y)^{4c -1} \,U(z)
\eeq
(see eq.(\ref{change variables})). Now, we have a related
Schrodinger equation defined in the $z$ variable
\beq %
\left[-\frac{d^2}{dz^2}+ \mathfrak{V}_4(z)\right]U(z)=m^2U(z),\label{sch_a4}
\eeq
with
\beq %
\mathfrak{V}_4(z) = \frac{1-4c}{(1-z^2)^2} \left[ 1+2(1-2c)z^2 \right]. \label{potential_a_4}
\eeq
We can see that the potential function diverges at $z=\pm1$ and so the boundary conditions
of this analog problem are $\{U(z=\pm1)=0,\ U'(z=\pm1)\ \rm finite\}$
which must be in order to match finite $u(y)$ solutions to
eq. (\ref{massive modes}) at $y\rightarrow\pm\infty$. %The value of $c$
%puts an obvious condition on the way $U(z(y))$ must behave at the boundary (for $c > 1/4$
%particular attention must be payed on its derivative).
After solving the quantum analog we have to transform back variables and functions
to check the finiteness and continuity of the original solution $u(y)$
in order to be physically acceptable.

We now better introduce the variable $\theta$ by means of
\beq z = \cos\,\theta \eeq
which results in equation
\beq %
U''(\theta) - \cot(\theta) \,\, U'(\theta) - \frac{1-4c}{\sin^2\theta}\left[1+2 (1- 2c)\cos^2\theta \right] U(\theta) = -m^2 \sin^2\!\theta \,U(\theta) \label{Utheta}
\eeq
for the analog wave function $U(\theta)$ with $U(\theta=\pi,0)=0$.

According to the arguments of localization seen in the previous section,
physically acceptable solutions require  $c > 0$ %\cite{CHRISTIANSEN}.
so we shall be restricted to that region.

\subsubsection{The c=1/4 case} %$\lambda = 30/34 \lambda_0$ > lambda_0 porque es negativo.
\begin{figure}[ht]
 \centering
\includegraphics[width=7.5cm,height=5.5cm]{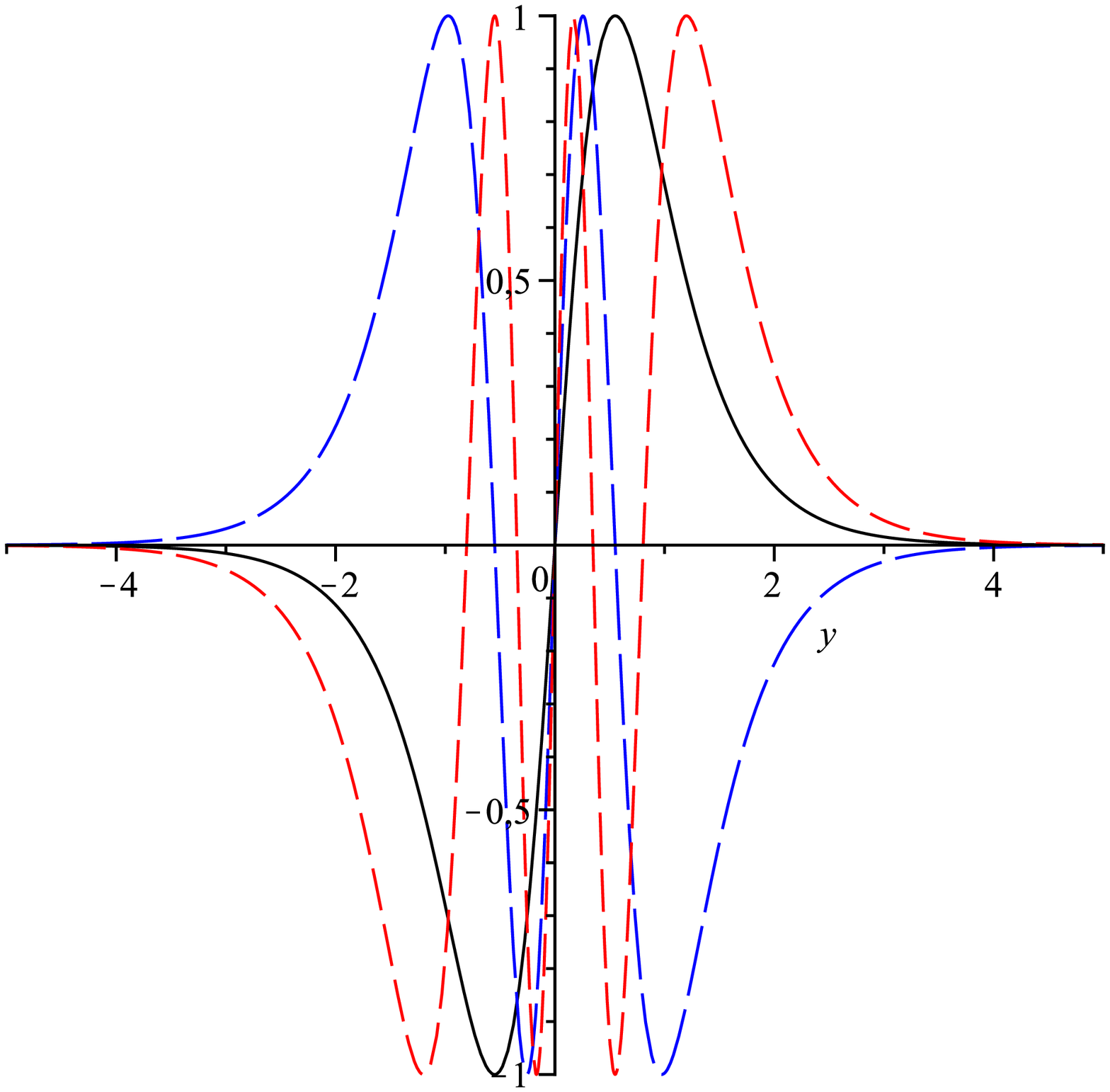}
\caption{\label{graf_a4_sch_c_1_4} Plot of $\sin (m \tanh(y))$, for $c=1/4$, $m=\pi$ (black line), $m=2\pi$ (long-dashed blue line), and $m=3\pi$ (dashed red line)).}
\includegraphics[width=7.5cm,height=5.5cm]{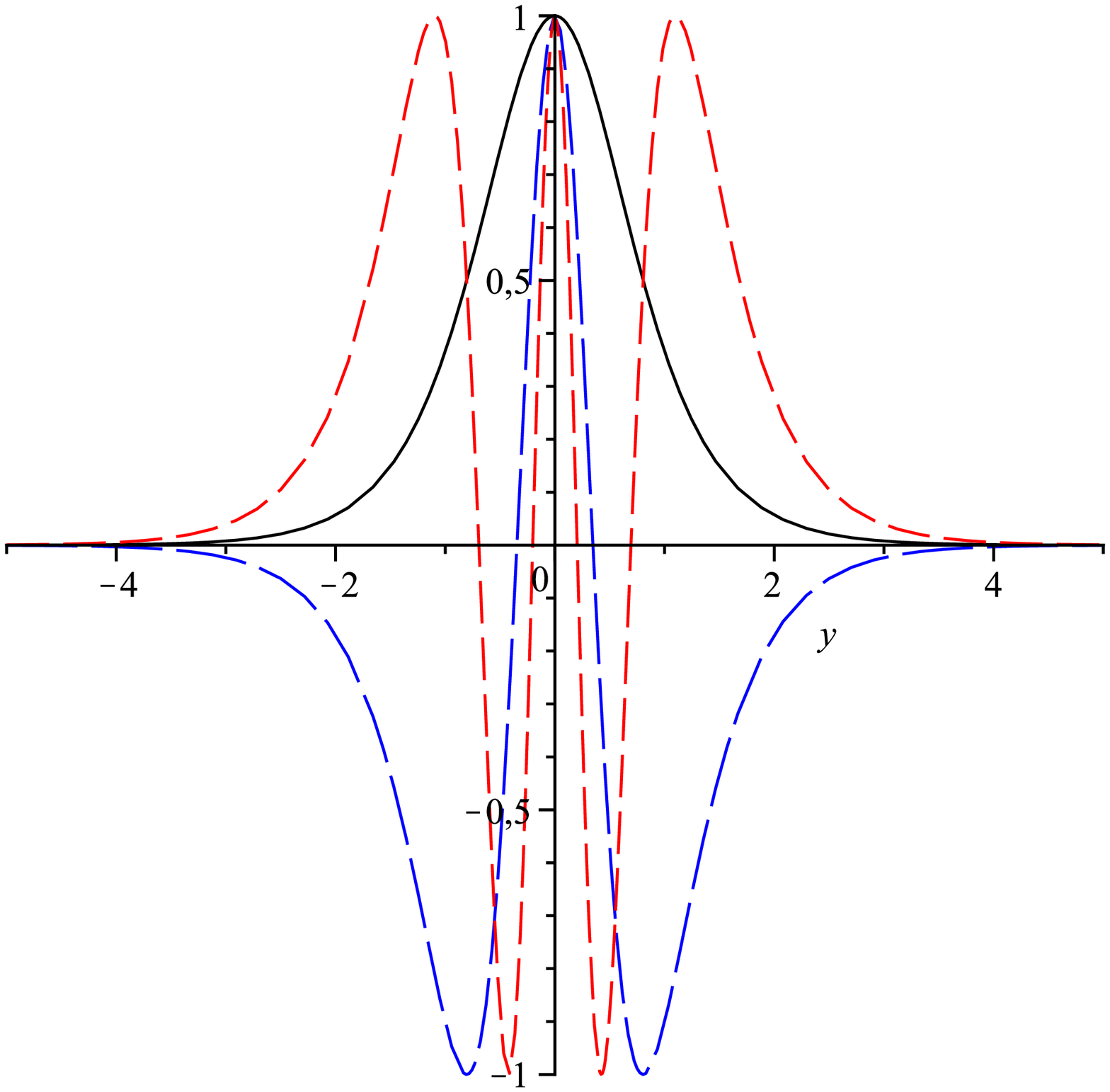}
\caption{\label{graf_a4_sch_cos_c_1_4} Plot of $\cos (m \tanh(y))$, for $c=1/4$, $m=\pi/2$ ( black line), $m=3\pi/2$ (long-dashed blue line), and $m=5\pi/2$ (dashed red line)).}
\end{figure}

Equation (\ref{Utheta}) gets strongly simplified for the value $c=1/4$. In this case we obtain
\beq
\frac{1}{\sin\theta} \frac{d}{d\theta} \left(\frac{1}{\sin\theta} U'(\theta) \right) +
m^2 \sin^2\!\theta ~U(\theta) = 0
\eeq
with solutions
\bea
U^{(1)}(\theta) = U_0\sin (m\cos(\theta))\\
U^{(2)}(\theta) = U_0\cos (m\cos(\theta)),
\eea
%
%\bea
%U^{(1)}(z) = U_0\sin (m z)\\
%U^{(2)}(z) = U_0\cos (m z)
%\eea as expected.
%
which in terms of the original variable and function read
\bea
u^{(1)}(y) = u_0\sin (m \tanh \!y) \label{Mt_c_1_4a} \\
u^{(2)}(y) = u_0\cos (m \tanh \!y) \label{Mt_c_1_4b},
\eea
(see Figs. \ref{graf_a4_sch_c_1_4}, \ref{graf_a4_sch_cos_c_1_4}). 
The zero-mode, $m=0$, is then related to $u(y)=u_0$ as already mentioned.
%
%%%%%%%%%%%%%%%%%%%%%%%%%%%%%%%%%%%%%%%%%%%%%%%%%%%%%%%%%%%%%%%%%%%%%%%%%%%%%%%%%%%
%%                                FIGURA
%%%%%%%%%%%%%%%%%%%%%%%%%%%%%%%%%%%%%%%%%%%%%%%%%%%%%%%%%%%%%%%%%%%%%%%%%%%%%%%%%%%
\smallskip

Since $\mathfrak{V}(z)$ is an even function (in this case trivial), solutions must have definite parity.
Besides, the potential divergence at $z= \pm 1$ implies that the corresponding solutions
are expected to be zero there.
Thus, antisymmetric solutions correspond to the eigenvalues of the Schrodinger equation (\ref{sch_a4}) $m= n \pi$ while symmetric solutions correspond to $m=(2n+1)\pi/2$,
with $n\in N$ (or simply $m=(n+1)\pi/2$ with $n$ even for symmetric solutions and odd for the antisymmetric ones).

\subsubsection{Other analytical solutions}% $\lambda> 29/34 \lambda_0$.
%For arbitrary values of $c$, $c\equiv(17+2\lambda\sqrt{3M^3})/4>0$, namely $\lambda> -\frac{17}{2\sqrt{3M^3}}=\lambda_0$.
If we perform the transformation
\beq
U(\theta) = \sin^\kappa \! \theta ~\cal{M}(\theta),
\eeq
in place of eq.(\ref{Utheta}) we get the following problem for $\cal{M}(\theta)$
\bea
\cal{M}''(\theta) + (2\kappa-1)\cot(\theta) \cal{M}'(\theta)+&& \left[ \kappa(\kappa-2) \cot^2\!(\theta) - \kappa + \frac{1-4c}{\sin^2\!\theta} [1+2(1-2c) \cos^2\!\theta] \right] \cal{M}(\theta) \nonumber \\
 &&~~~~~~~~~~~~~~~~~~~~~~~~~~~~~~~~~~~~~~~~~~~~= -m^2 \sin^2\!\theta \cal{M}(\theta).
\eea%
%
%%%%%%%%%%%%%%%%%%%%%%%%%%%%%%%%%%%%%%%%%%%%%%%%%%%%%%%%%%%%%%%%%%%%%%%%%%%%%%%%%%%%%%%%%%%%%%
%                    TABELA PARA VALORES DE m PARA O CASO c=5/8 c~=-1
%%%%%%%%%%%%%%%%%%%%%%%%%%%%%%%%%%%%%%%%%%%%%%%%%%%%%%%%%%%%%%%%%%%%%%%%%%%%%%%%%%%%%%%%%%%%%%
\begin{table}[hb]
\caption{\label{table2} List of the first 21 values of $m_s$ (symmetric solutions) and $m_a$ (antisymmetric ones) for $c=5/8$ }
\vskip 0.2cm%
\begin{tabular}{c c c }
  \hline  \hline
\,   \, & $m_s$\, & \, $m_a$\\ \hline
 \,   \, & \, 0.0000000 \, & \, $--$ \\
  \,   \, & \,4.0649860  \, & \, 2.3807959 \\  %2.380795874
   \,   \, & \,7.2962115  \, & \, 5.6914019 \\  %5.691401880
    \,   \, & \,10.4780880 \, & \, 8.8902613\\   %8.890261250
     \,   \, & \,13.6431458 \, & \, 12.0619596\\  %12.06195958
      \,   \, & \,16.8002756 \, & \, 15.2224185\\  %15.22241845
       \,   \, & \,19.95298930\, & \, 18.3770538\\  %18.37705380
        \,   \, & \,23.1029720 \, & \, 21.5282531\\  %21.5282531
         \,  \, & \,26.2511409 \, & \, 24.67724335\\  %24.67724335
          \,   \, & \,29.3980716 \, & \, 27.8247238\\  %27.82472376
           \,  \, & \,32.5440439 \, & \, 30.97112352\\  % 30.97112352
                      \,  \, & \,$\dots$ \, & \,$\dots$  \\
            \hline
             \hline
\end{tabular}
\end{table}%                         END TABELA a=4, c=5/8
Now, we can choose a convenient power for the last transformation, $\kappa=1/2$,
in order to turn this into
\beq
\cal{M}''(\theta) - 4 \left[ \cot^2(\theta) \left(4c^2-4c+\frac{15}{16} \right) +\frac{3}{8} -c \right]  \cal{M}(\theta)=-m^2 \sin^2\!\theta \cal{M}(\theta)
\eeq
which, for $4c^2-4c+\frac{15}{16}=0$, is known as the Mathieu differential equation
\beq
M''(\theta) + \left( 4c-3/2 +m^2 \sin^2 \theta \right) M(\theta)=0,  \label{mathieu}
\eeq
with $c=3/8$ or $c=5/8$.

\subsubsection{The case $c=5/8$}

In this case, Eq. (\ref{mathieu}) results in
\beq
M''(\theta)+\left(1+\frac{m^2}{2}-\frac{m^2}{2} \cos(2\theta) \right) M(\theta) = 0
\eeq
%
%%%%%%%%%%%%%%%%%%%%%%%%%%%%%%%%%%%%%%%%%%%%%%%%%%%%%%%%%%%%%%%%%%%%%%%%%%%%%%%%%%%
%%                        FIGURA em z para c = 5/8
%%%%%%%%%%%%%%%%%%%%%%%%%%%%%%%%%%%%%%%%%%%%%%%%%%%%%%%%%%%%%%%%%%%%%%%%%%%%%%%%%%%
\smallskip
\begin{figure}[ht]
 \centering
\includegraphics[width=7.5cm,height=5.5cm]{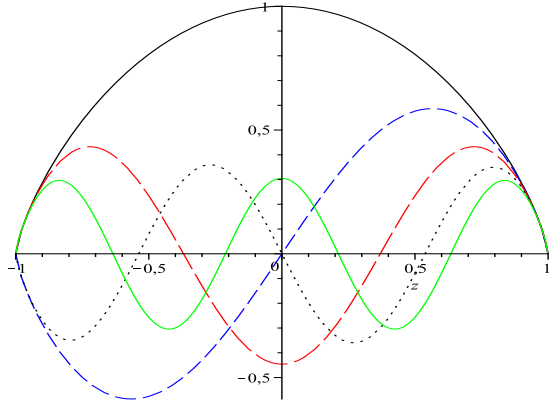}
\caption{\label{graf_c_5_8} Symmetric and antisymmetric eigenfunctions $U$ in $z$ space for
$c = 5/8$; $m=0$ (black line, symmetric), $m=2.380795874$ (dashed blue line, anti-symmetric), $m=4.0649860$ (long-dashed red line, symmetric), $m=5.6914019$ (dotted black line, antisymmetric), and $m=7.2962115$ (solid green line, symmetric).}
\end{figure}
whose analytic solutions are the general Mathieu functions
\bea
M^{(1)}(\theta)=Mc\left(1+\frac{m^2}{2},\frac{m^2}{4}, \theta\right)\\
M^{(2)}(\theta)=Ms\left(1+\frac{m^2}{2},\frac{m^2}{4}, \theta\right),
\eea
which, in terms of the $z$ variable,  result in the analog wave-functions  $U(z)$
\bea
U^{(1)}(z) = (1-z^2)^{1/4} Mc \left(1+\frac{m^2}{2},\frac{m^2}{4}, \arccos (z)\right) \label{Mt_c_5_8a} \\
U^{(2)}(z) = (1-z^2)^{1/4} Ms \left(1+\frac{m^2}{2},\frac{m^2}{4}, \arccos(z)\right) \label{Mt_c_5_8b}
\eea
(see \cite{ERDELYI} for details about Mathieu functions).

As mentioned above, the boundary conditions of the present problem are
$\{U(z=\pm1)=0,\ U'(z=\pm1)\ \rm finite\}$,
related to finite $u(y)$ solutions to the original equation, recalling that $y\in(-\infty, \infty)$.
The first set of solutions, $U^{(1)}(z)$, is not physically interesting because
the derivatives of these functions are divergent at the boundary.
The reason is that $Mc(\arccos (z))$ cannot be zero at $z=1$ for any value of $m$.
The second set, on the other hand, has physically
acceptable solutions for a discrete set of values of $m$, the (twenty) first of which we show
in Table \ref{table2}.  These solutions are symmetric or antisymmetric, as expected (see
Fig. \ref{graf_c_5_8}). Note the presence of a zero mode.
%%%%%%%%%%%%%%%%%%%%%%%%%%%%%%%%%%%%%%%%%%%%%%%%%%%%%%%%%%%%%%%%%%%%%%%%%%%%%%%%%%%%%%%
% 					FIGURA  c = 5/8 -> c~ = -1
%%%%%%%%%%%%%%%%%%%%%%%%%%%%%%%%%%%%%%%%%%%%%%%%%%%%%%%%%%%%%%%%%%%%%%%%%%%%%%%%%%%%%%%
\smallskip
\begin{figure}[ht]
 \centering
    \includegraphics[width=7.5cm,height=5.5cm]{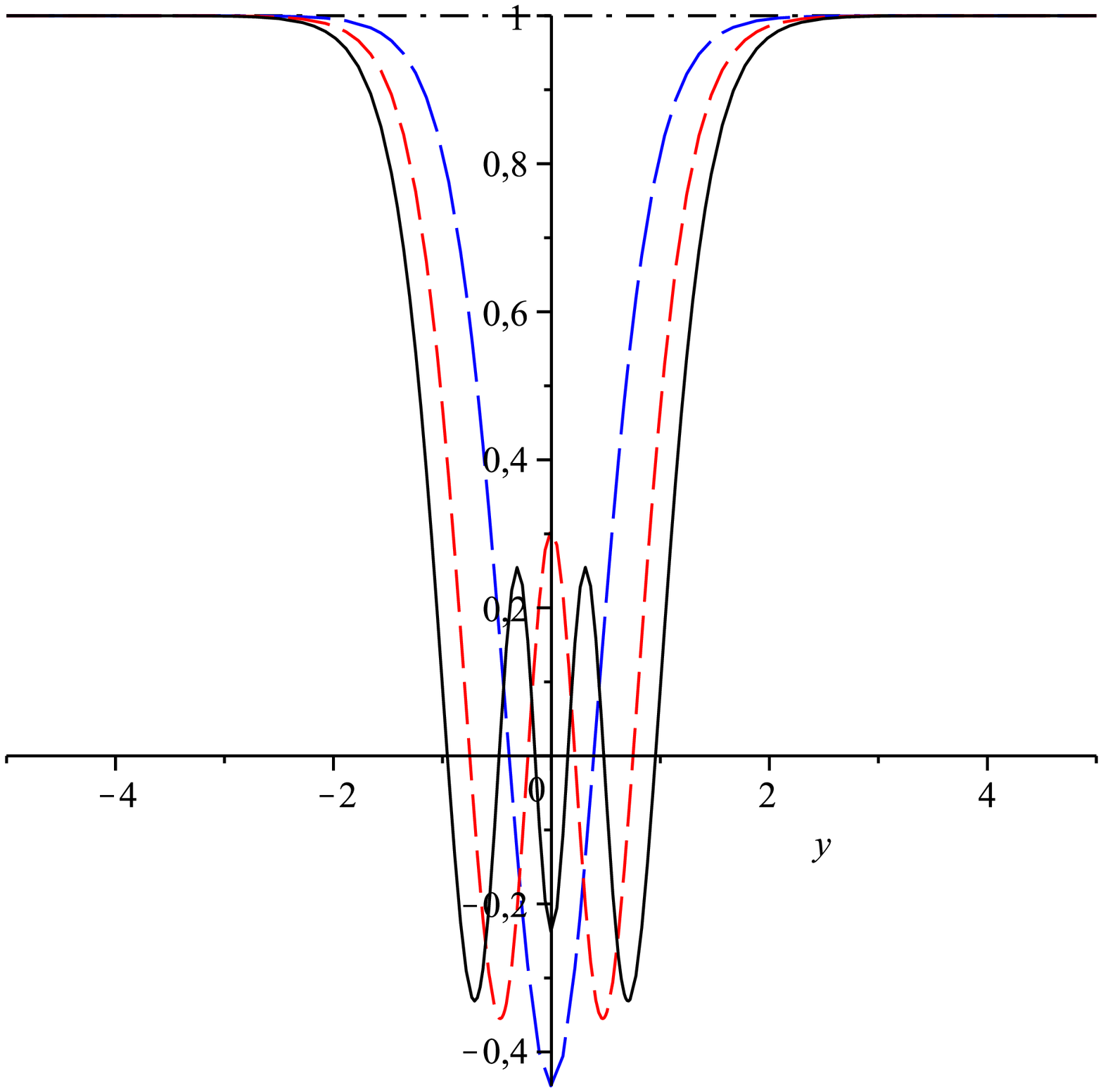}
\caption{\label{graf_c_5_8_sim_y} Symmetric solutions $u^{(2)}(y)$ (Eq. (\ref{u_y_c_5_8_})) for
$c = 5/8$; $m=0$ (dash-dotted black line), $m=4.0649860$ (long-dashed blue line), $m=7.2962115$ (
dashed red line), and $m=10.4780880$ (solid black line).}
\includegraphics[width=7.5cm,height=5.5cm]{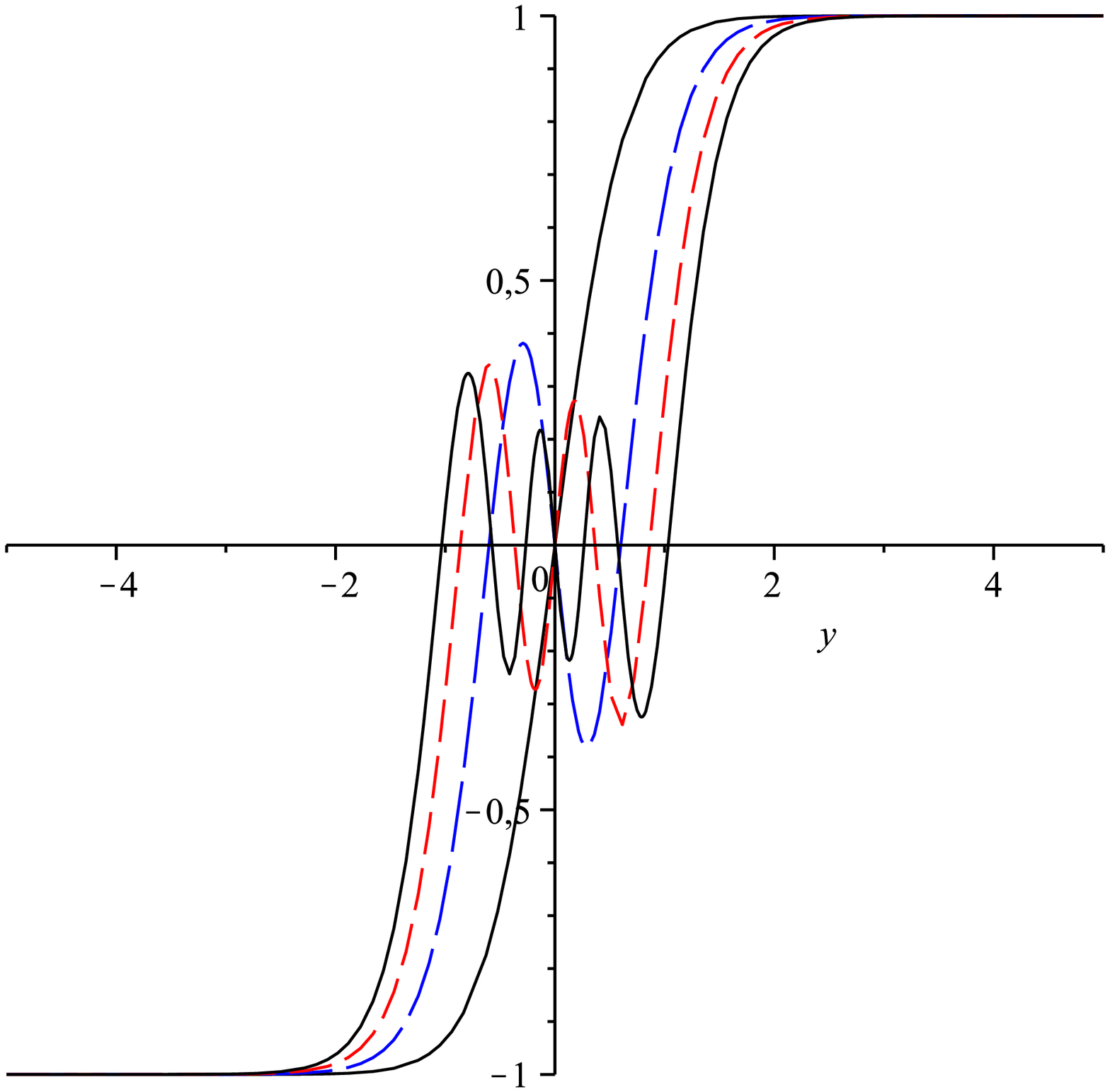}
\caption{\label{graf_c_5_8_asim_y} Antisymmetric solutions $u^{(2)}(y)$ for $c = 5/8$; $m=2.3807959 $ (solid black line), $m=5.6914019$ (long-dashed blue line),  $m=8.8902613$ (dashed red line), and $m=12.0619596$ (solid black line).}
\end{figure}
%
%%%%%%%%%%%%%%%%%%%%%%%%%%%%%%%%%%%%%%%%%%%%%%%%%%%%%%
%%                    TABLE   c = 3/8
%%%%%%%%%%%%%%%%%%%%%%%%%%%%%%%%%%%%%%%%%%%%%%%%%%%%%%
\begin{table}[hb]
\caption{\label{table3} List of the first 20 values of $m_s$ and $m_a$ for ${c}=3/8$.}
\vskip 0.2cm%
\begin{tabular}{c c c }
  \hline  \hline
\,  \, & $m_s$\, & \, $m_a$\\ \hline
 \,  \, & \, $--$ \, & \, $--$ \\
  \,  \, & \,1.14718042  \, & \, 2.72632477 \\  %1.1471804178  %2.726324772
   \,  \, & \,4.30206964  \, & \, 5.87592007 \\  %4.30206964  %5.875920066
    \,  \, & \,7.44879288  \, & \, 9.02110025 \\  %7.448792878  %9.021100246
     \,  \, & \,10.59305044 \, & \,12.16475984 \\  %10.59305044 %12.16475984
      \,  \, & \,13.73629872 \, & \,15.30771212 \\  %13.73629872 %15.30771212
       \,  \, & \,16.87903028 \, & \,18.45027434 \\  %16.87903028 %18.4502743412
        \,  \, & \,20.02145951 \, & \,21.59259704 \\  %20.02145951 %21.59259704
         \,  \, & \,23.16369547 \, & \,24.65270800 \\  %23.16369547 %24.652708041
          \,  \, & \,26.30579984 \, & \,27.87681505 \\  %26.30579984 % 27.87681505
           \,  \, & \,29.44781028 \, & \,31.01878823 \\  %27.278215069 % 25.965544530
           \,  \, & \,$\dots$ \, & \,$\dots$  \\  %27.278215069 % 25.965544530
            \hline
             \hline
\end{tabular}
\end{table} %        END TABLE  c = 3/8

In terms of $y$, we have
\bea
u^{(1)}(y) = \cosh \!y ~Mc \left(1+\frac{m^2}{2},\frac{m^2}{4}, \arccos (\tanh \!y)\right) \\
u^{(2)}(y) = \cosh \!y ~Ms \left(1+\frac{m^2}{2},\frac{m^2}{4}, \arccos (\tanh \!y)\right), \label{u_y_c_5_8_}
\eea
where the set $u^{(1)}(y)$ diverges when $y \rightarrow \pm \infty$, as due from the
comments above, so we just keep the solutions
$u^{(2)}(y)$ (see Figs. \ref{graf_c_5_8_sim_y} and \ref{graf_c_5_8_asim_y}).
%
%
%
%%%%%%%%%%%%%%%%%%%%%%%%%%%%%%%%%%%%%%%%%%%%%%%%%%%%%%%%%%%%%%%%%%%%%%%%%%
%%				 							CASE c = 3/8 c~=1
%%%%%%%%%%%%%%%%%%%%%%%%%%%%%%%%%%%%%%%%%%%%%%%%%%%%%%%%%%%%%%%%%%%%%%%%%%
\subsubsection{The case $c=3/8$}
%      $  \lambda> 31/34 \lambda_0    $.
%
Now, Eq. (\ref{mathieu}) reads
\beq
M''(\theta)+\left(\frac{m^2}{2}-\frac{m^2}{2} \cos(2\theta) \right) M(\theta) = 0
\label{mathieu3/8}
\eeq
with solutions given by
\bea
M^{(1)}(\theta)=Mc\left(\frac{m^2}{2},\frac{m^2}{4}, \theta\right)\\
M^{(2)}(\theta)=Ms\left(\frac{m^2}{2},\frac{m^2}{4}, \theta\right),
\eea
corresponding to
\bea
U^{(1)}(z) = (1-z^2)^{1/4} Mc \left(\frac{m^2}{2},\frac{m^2}{4}, \arccos (z)\right) \label{Mt_c_3_8a}\\
U^{(2)}(z) = (1-z^2)^{1/4} Ms \left(\frac{m^2}{2},\frac{m^2}{4}, \arccos(z)\right). \label{Mt_c_3_8b}
\eea
in the $z$ space with the boundary conditions already seen.
%
%%%%%%%%%%%%%%%%%%%%%%%%%%%%%%%%%%%%%%%%%%%%%%%%%%%%%%%%%%%%%%%%%%%%%%%%%%%%%%%%%%%%%%%
% 								FIGURAS  c = 3/8 -> c~ = 1  em (z)
%%%%%%%%%%%%%%%%%%%%%%%%%%%%%%%%%%%%%%%%%%%%%%%%%%%%%%%%%%%%%%%%%%%%%%%%%%%%%%%%%%%%%%%
\smallskip
\begin{figure}[ht]
\centering
\includegraphics[width=7.5cm,height=5.5cm]{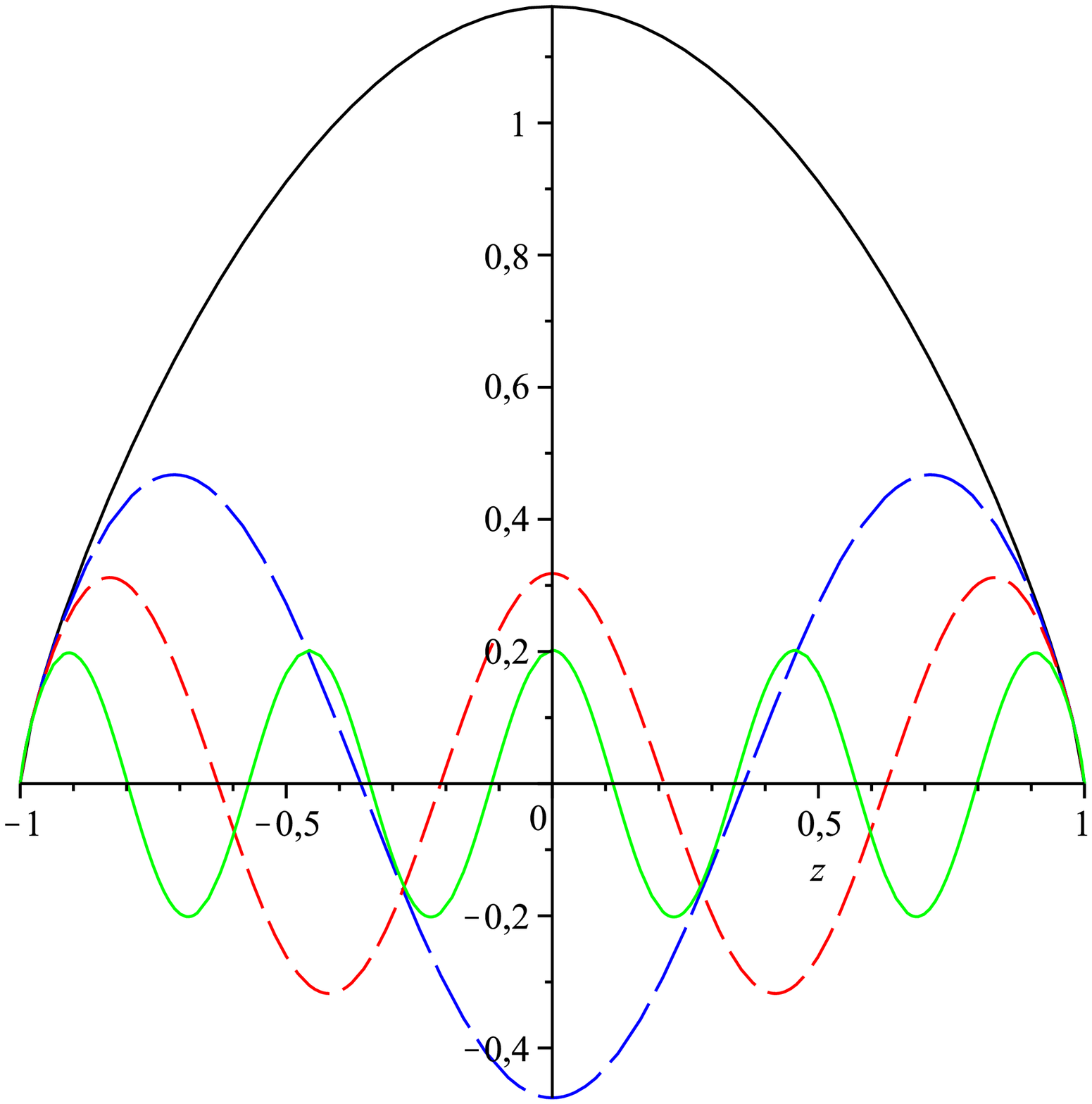}
\caption{\label{graf_sim_c_3_8} Symmetric solutions $U^{(2)}(z)$  for
$c = 3/8$; $m=1.14718042$ (solid black line), $m=4.30206964$ (long-dashed blue line), $m=7.44879288$ (dashed red line), and $m=13.73629872$ (solid green line).}
%\subfigure[Massive modes ($m^2=1,10$)]
\includegraphics[width=7.5cm,height=5.5cm]{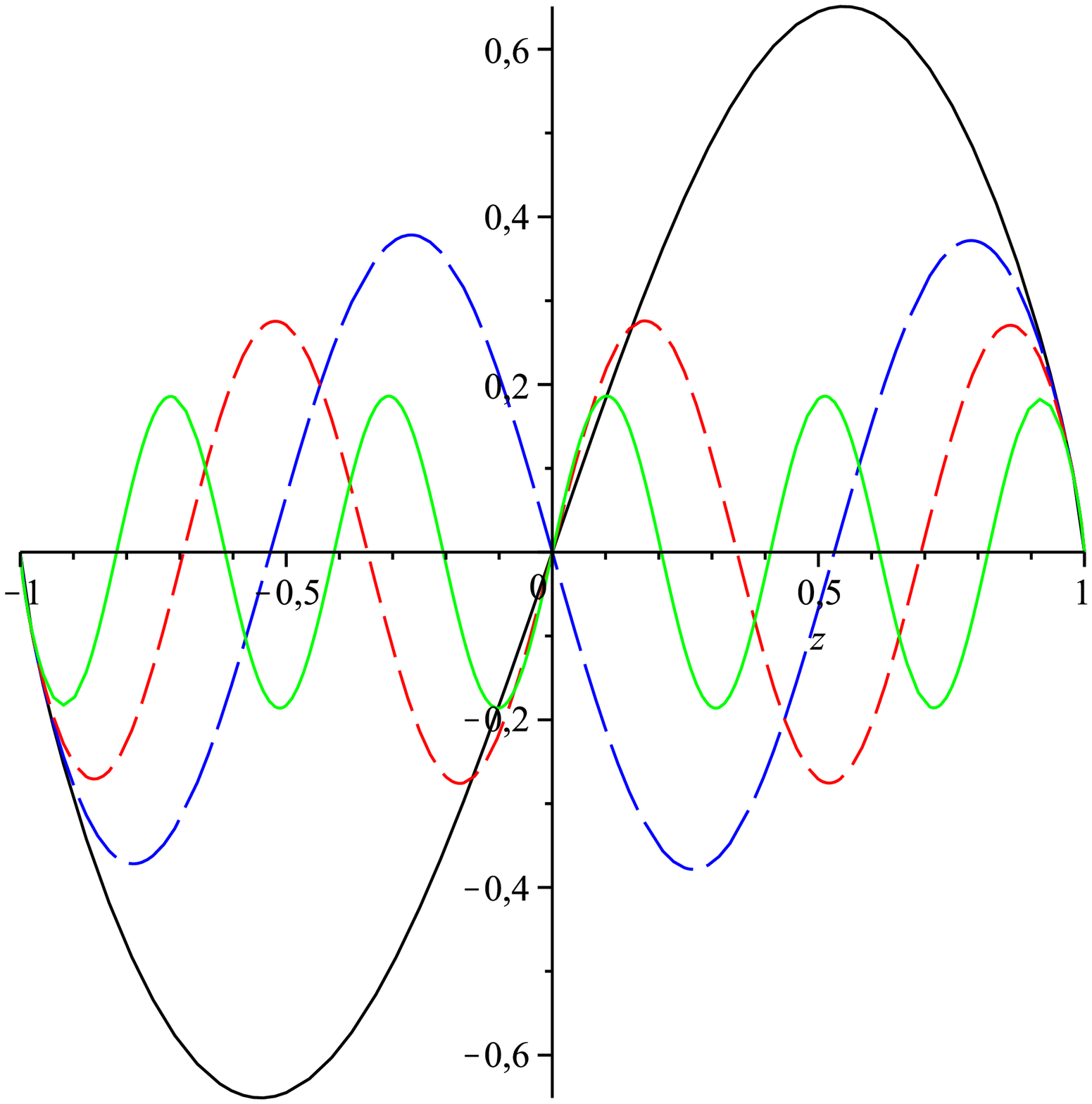}
\caption{\label{graf_asim_c_3_8} Antisymmetric solutions $U^{(2)}(z)$ for $c = 3/8$; $m=2.72632477$ (solid black line), $m=5.87592007$ (long-dashed blue line),  $m=9.02110025$ (dashed red line), and $m=15.30771212$ (solid green line).}
\end{figure}
\begin{figure}[ht]
\centering
\includegraphics[width=7.5cm,height=5.5cm]{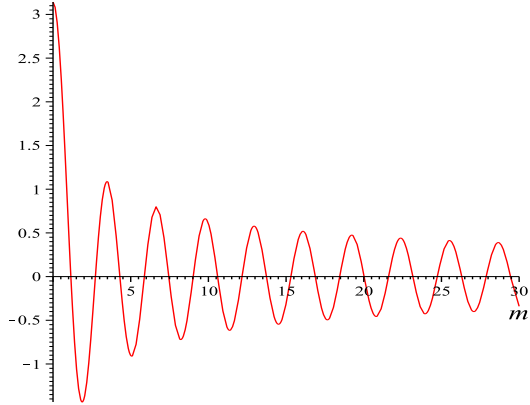}
\caption{\label{zerosMathieu} Mathieu function at the boundary ($\theta=\pi$) as a function of the mass for $c = 3/8$, $Ms\left(\frac{m^2}{2},\frac{m^2}{4}, \theta=\pi \right)$. At the other boundary, $Ms\left(\frac{m^2}{2},\frac{m^2}{4}, \theta=0 \right)=0$.}
\end{figure}
%%%%%%%%%%%%%%%%%%%%%%%%%%%%%%%%%%%%%%%%%%%%%%%%%%%%%%%%%%%%%%%%%%%%%%%%%%%%%%%%%%%%%%%

As we discussed in the previous ($c=5/8$) case, only the second set of solutions is physically relevant and just for a discrete (infinite) sequence  of  $m$ eigenvalues. For such values
solutions have definite parity according to $\mathfrak{V}(z)$, Eq. (\ref{potential_a_4}),
(see Figs. (\ref{graf_sim_c_3_8}) and (\ref{graf_asim_c_3_8})).

Note that the solutions to the Schrodinger equation (the analytical expressions (\ref{Mt_c_3_8a}) and (\ref{Mt_c_3_8b}))
are not compatible with a zero-mode for the $z$-boundary conditions given above. Actually, as a general result, for any value excluded from the sequence starting in Table \ref{table3}, $U^{(1)}(z)$ -eq. (\ref{Mt_c_3_8a})- has divergent derivatives at $z=\pm 1$ and $U^{(2)}(z)$ -eq. (\ref{Mt_c_3_8b})- is not even symmetric. For this reason the zero-mass solutions of eq. (\ref{mathieu3/8}), $M(\theta, m=0)\in\{\rm cons., \theta\}$, do not correspond to valid solutions of the Schrodinger problem.
In Fig.\ref{zerosMathieu} we can see all the first mass values of the sequence which nullify the Mathieu functions $Ms$ at the boundary. These values, also listed in Table \ref{table3}, guarantee finite derivatives of  $U^{(2)}(z)$. The absence of the zero mode in this list indicates a limitation of the Schrodinger analogue approach. We will come again to this point in the next Section.
%%%%%%%%%%%%%%%%%%%%%%%%%%%%%%%%%%%%%%%%%%%%%%%%%%%%%%%%%%%%%%%%%%%%%%%%%%%%%%%%%%%%%%%
% 								FIGURAS  c = 3/8 -> c~ = 1    em  (y)
%%%%%%%%%%%%%%%%%%%%%%%%%%%%%%%%%%%%%%%%%%%%%%%%%%%%%%%%%%%%%%%%%%%%%%%%%%%%%%%%%%%%%%%
\begin{figure}[hb]
 \centering
    \includegraphics[width=7.5cm,height=5.5cm]{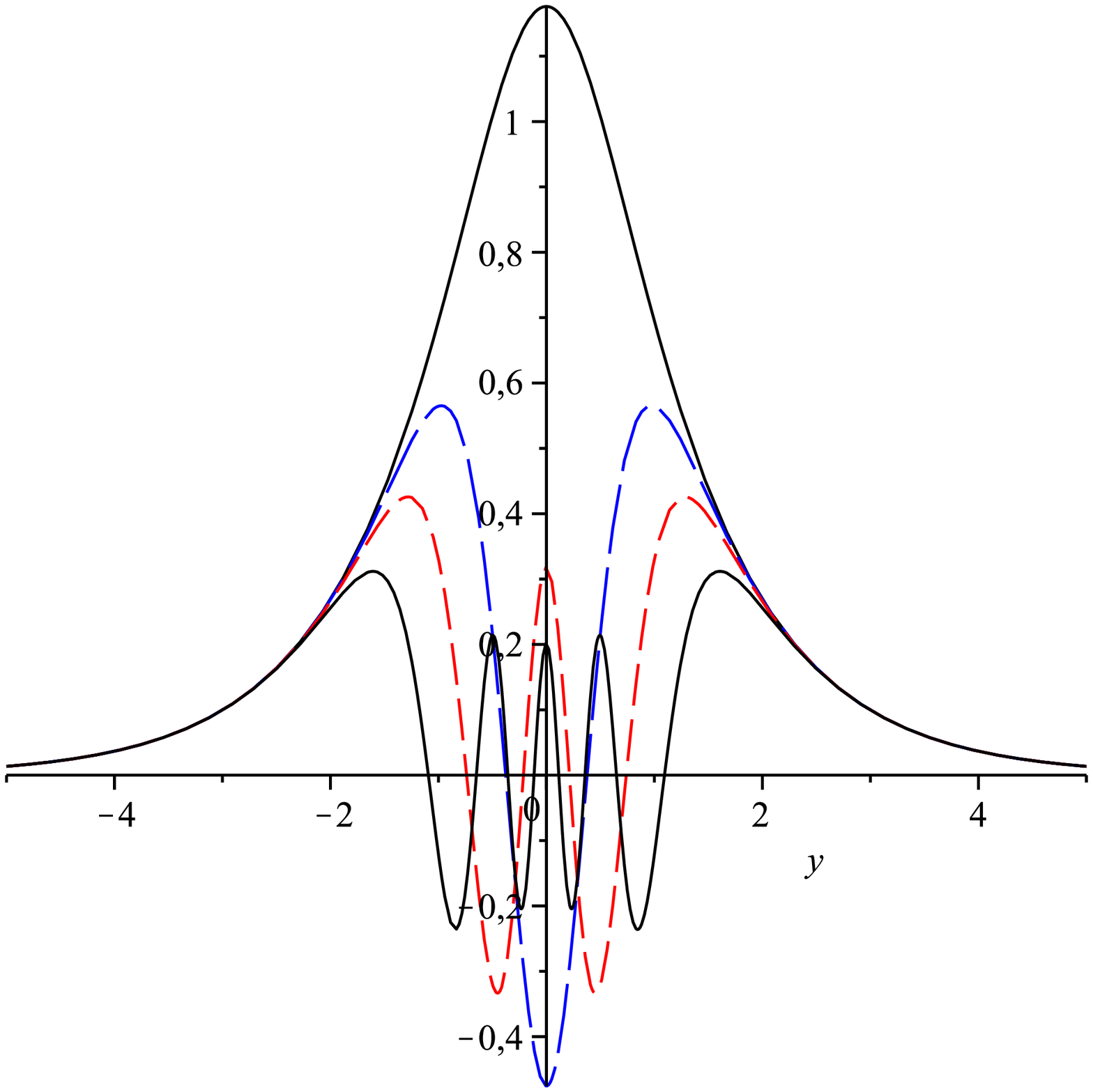}
\caption{\label{graf_c_3_8_sim_y} Symmetric solutions $u^{(2)}(y)$, eq.(\ref{u_y_c_3_8}),  for
$c = 3/8$; $m=1.14718042$ (solid black line), $m=4.30206964$ (long-dashed blue line), $m=7.44879288$ (dashed red line), and $m=13.73629872$ (solid black line).}
\includegraphics[width=7.5cm,height=5.5cm]{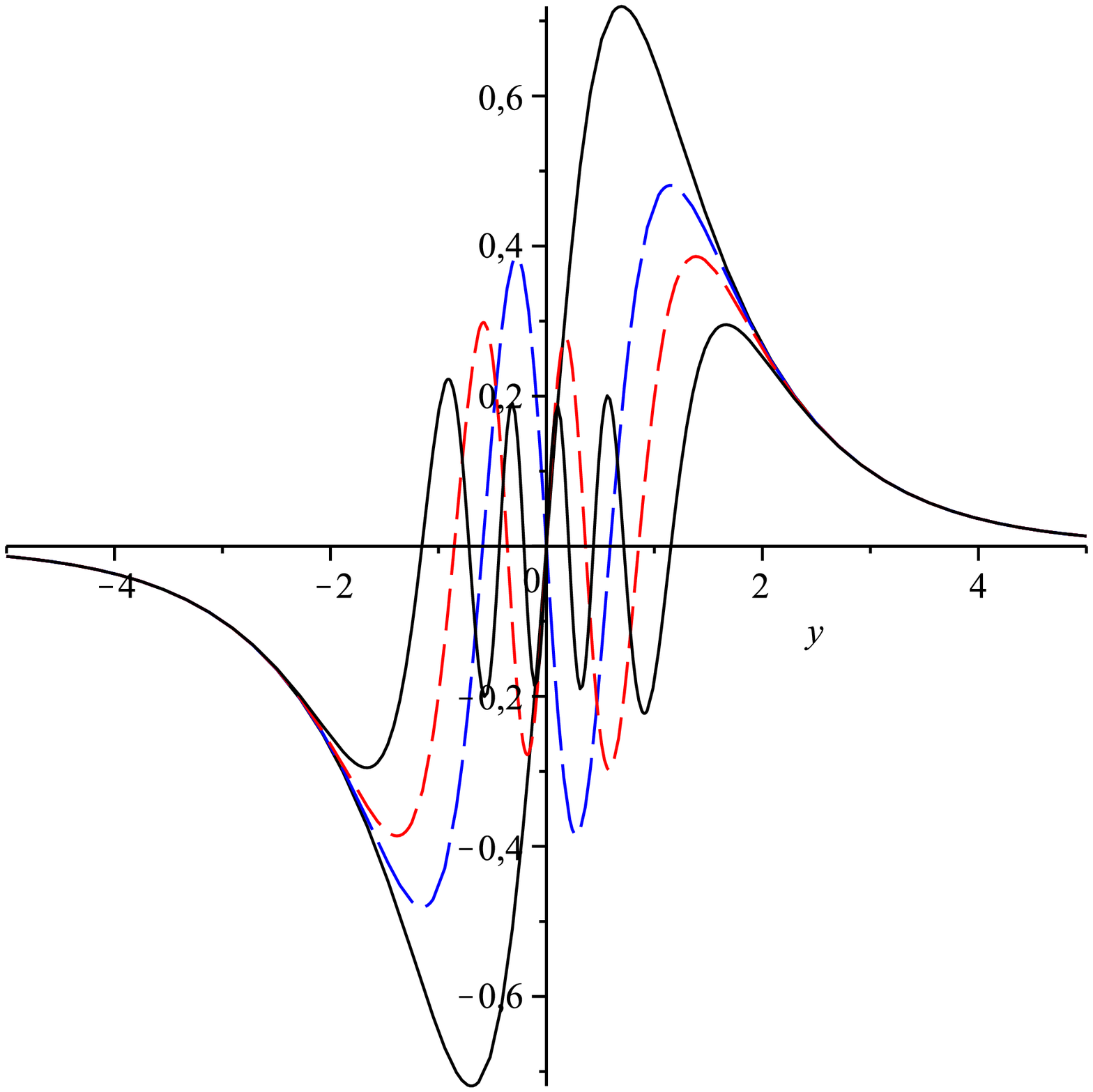}
\caption{\label{graf_c_3_8_asim_y} Antisymmetric solutions $u^{(2)}(y)$, eq.(\ref{u_y_c_3_8}), for $c  = 3/8$; $m=2.72632477 $ (solid black line), $m=5.87592007$ (long-dashed blue line),  $m=9.02110025$ (dashed red line), and $m=15.30771212$ (solid black line).}
\end{figure}
%%%%%%%%%%%%%%%%%%%%%%%%%%%%%%%%%%%%%%%%%%%%%%%%%%%%%%%%%%%%%%%%%%%%%%%%%%%%%%%%%%%%%%%
In terms of $y$ we have
\bea
u^{(1)}(y) = Mc \left(\frac{m^2}{2},\frac{m^2}{4}, \arccos (\tanh \!y)\right) \label{u_y_c_3_8a}\\
u^{(2)}(y) = Ms \left(\frac{m^2}{2},\frac{m^2}{4}, \arccos (\tanh \!y)\right), \label{u_y_c_3_8}
\eea
of which $u^{(2)}$ represents the only non-divergent set of solutions, as we illustrate in Figs. (\ref{graf_c_3_8_sim_y}) and (\ref{graf_c_3_8_asim_y}) for the first quantum values of $m$.

\section{The Confluent Heun equation \label{heun}}
We now investigate our original problem by relaxing the quantum analog condition.
In order to obtain the general solution of Eq. (\ref{massive modes}) we perform the
following change of variable
\beq x = \tanh \!y. \eeq
It  maps the $y$ space to $x\in(-1,1)$ and we will eventually transform it back
in order to come into the original space and variables.
Note that Eq. (\ref{massive modes}) is symmetric under a parity transformation and thus
the differential equation
admits even as well as odd parity solutions, as it should.
Now,  Eq. (\ref{massive modes}) becomes

%%%%%%%%%%%%%%%%%%%%%%%%%%%%%%%%% EQ PRINCIPAL %%%%%%%%%%%%%%%%%%%%%%%%%%
\beq
u''(x) + (2- \tilde{c}) \frac{x}{x^2-1} u'(x) + m^2 (1-x^2)^{\frac{a}{2} -2} u(x) = 0.
\label{eqUa4}
\eeq
which is an homogeneous second-order linear differential equation with polynomial
coefficients provided $a$ is even.  Here $\tilde{c}= a(1-2c)$ so that
 $\tilde{c}= (-\infty ,a)$. Now, Eq.(\ref{eqUa4}) looks
more familiar if we change $x^2$ into $z$

\beq
u''(z) + \left( \frac{1/2}{z}+ \frac{1-\tilde{c}/2}{z-1} \right) u'(z)
+ \frac{m^2}{4} \frac{u(z)}{z} = 0. \label{eqUa4b}
\eeq
for the case under study. %, $a=4$,
Formally, this equation has two regular singular (Fuchsian) points  at $z=0, 1$,
and an irregular one at $z=\infty$.
%(although the original region of interest be restricted to $z<1$).
%
This is known as a Confluent Heun equation \cite{HEUN, heunDE, HOUNKONNOU}.

We can compare Eq. (\ref{eqUa4b}) with the canonical non-symmetrical
general form of the  confluent Heun equation as
given in \cite{HOUNKONNOU, heunDE, CLEB-TERRAB}
\bea
Hc''(z) &+& \left(\alpha + \frac{\beta + 1}{z} + \frac{\gamma + 1}{z-1} \right)Hc'(z)  \nonumber \\
&+& \left[ \frac{[\delta + \frac{\alpha}{2}(\beta + \gamma +2)]z + \eta + \frac{\beta}{2}  + \frac{1}{2} (\gamma - \alpha)(\beta+1)}{z(z-1)} \right] Hc(z) = 0, \label{heunc}
\eea
whose solutions around $z=0$ are denoted by
\bea
H^{(1)}&=& Hc(\alpha,\beta,\gamma,\delta,\eta; z) \\
H^{(2)}&=& z^{-\beta}\,Hc(\alpha,-\beta ,\gamma , \delta, \eta; z).%
\eea
In general, there are two linearly independent local series solutions around
each singular point. In the region of interest,  $z<1$, we look for
a regular local solution around $z=0$ which is defined  by the Heun series as
\beq
Hc(z) = \sum_{n=0}^\infty d_n \, z^n.
\eeq
Here the constants $d_n$ (with $d_{-1}=0$ and $d_0=1$)
are determined by the three-term recurrence relation \cite{FIZIEV}
\bea
A_n d_n = B_n d_{n-1}+C_n d_{n-2}, \label{recurrence}
\eea
where
\bea
A_n &=& 1+\frac{\beta}{n} \rightarrow 1- \frac{1}{2n} \\
B_n &=& 1+ \frac{-\alpha + \beta + \gamma-1}{n}+\frac{\eta+(\alpha-\beta-\gamma)/2-\alpha \beta/2+\beta \gamma/2}{n^2} \nonumber \\
&\rightarrow & 1+ \frac{-\tilde{c}/2 -3/2}{n}+\frac{\tilde{c}/2+1/2-m^2/4}{n^2} \\
C_n &=& \frac{1}{n^2} \left(\delta+\frac{\alpha(\beta+\gamma)}{2}+\alpha (n-1) \right) \rightarrow \frac{m^2}{4n^2}.
\eea

By comparing Eqs. (\ref{eqUa4b}) and (\ref{heunc}), it is easy to
identify $\alpha=0$, $\beta=-1/2$, $\gamma =-\tilde{c}/2$, $\delta=m^2/4$, and $\eta=\tilde{c}/8 +1/4-m^2/4$.
Then the solutions of Eq. (\ref{massive modes}) are given by
\bea
u^{(1)}(y) &=&Hc \left( 0,-\frac{1}{2},-\frac{\tilde{c}}{2},\frac{m^2}{4},\frac{1}{4}+\frac{\tilde{c}}{8}-\frac{m^2}{4}; \, \tanh^2 \!y \right) \label{hc_a4_s}\\
u^{(2)}(y)
&=& \tanh\!y\, Hc \left( 0,\frac{1}{2},-\frac{\tilde{c}}{2},\frac{m^2}{4},\frac{1}{4}+\frac{\tilde{c}}{8}-\frac{m^2}{4}; \, \tanh^2 \!y \right) \label{hc_a4_a}
\eea
for arbitrary values of $\tilde{c}$ (or $c$), namely of the dilaton coupling constant. The
conditions these Heun $u(y)$ solutions must obey to be acceptable are the original ones, i.e.
finiteness and continuity in the whole space.

%%%%%%%%%%%%%%%%%%%%%%%%%%%%%%%%%%%%%%%%%%%%%%%%%%%%%%
%%                    TABLE   ~c = -30      lambda=0
%%%%%%%%%%%%%%%%%%%%%%%%%%%%%%%%%%%%%%%%%%%%%%%%%%%%%%
\begin{table}[ht]
\caption{\label{table4} List of first values of $m_s$ and $m_a$ for $\tilde{c}=-30$.}
\vskip 0.2cm%
\begin{tabular}{c c c }
  \hline  \hline
\,  \, & $m_s$\, & \, $m_a$\\ \hline
 \,  \, & \, $0$ \, & \, $-- $ \\
  \,  \, & \,8.69355330  \,& \, 5.90953031    \\
   \,  \, & \,13.16860126  \,& \, 11.03547247 \\
    \,  \, & \,17.10643340  \,& \, 15.17836781 \\
     \,  \, & \,20.80307154 \, & \, 18.97642242  \\
      \,  \, & \,24.36269818 \, & \, 22.59620420  \\
      \,  \, & \,$\dots$ \, & \, $\dots$ \\
            \hline
             \hline
\end{tabular}
\end{table} %        END TABLE
%
%m0 = 0 (dash dotted)
%m1=8.69355330552734439468009280033 (blue dashed)
%m2=13.16860126522520821312689064335 (red dashed)
%m3=17.106433403232751139375859221 (black )
%m4=20.803071549889780037679544048 (brown )
%m5=24.36269818820578194292247443 (orange)
%ma1 = 5.9095303113914788030417271186406 (black,solid)
%ma2 = 11.03547247660073165891054433317 (blue, dashed)
%ma3 = 15.1783678147615513637498457334 (red, dashed)
%ma4 = 18.9764224218760696324470672012 (brown, solid)
%ma5 = 22.596204203543992213822522858

\begin{figure}[ht]%''NORMALIZADAS'' segun criterio de area absoluta con respecto a la asintota
\centering
\includegraphics [height=4.5cm]{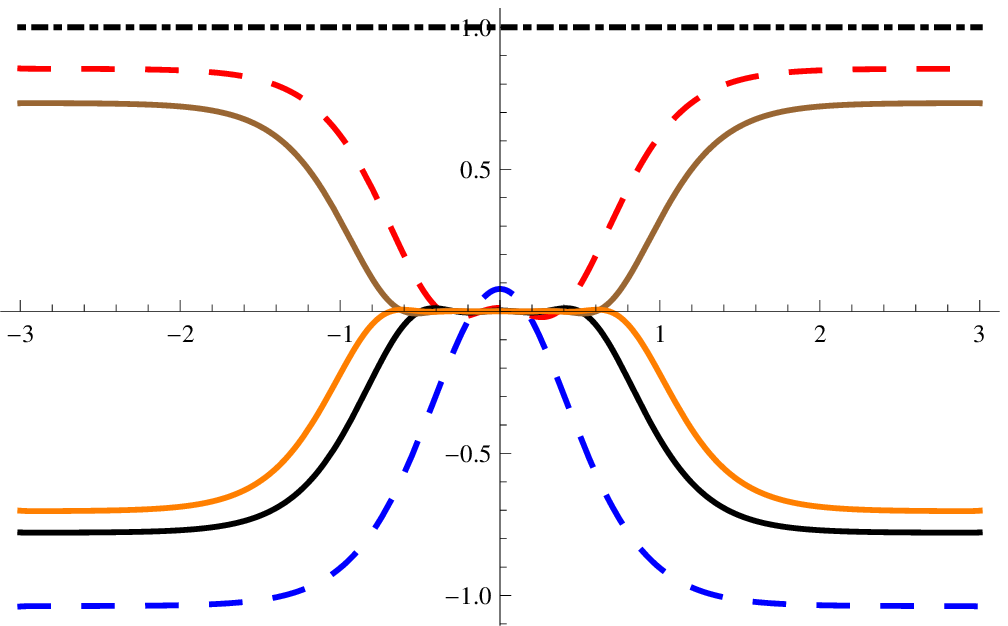}
\hspace{0.5cm}\subfigure[~]{
\includegraphics[width=4cm,height=3.0cm]{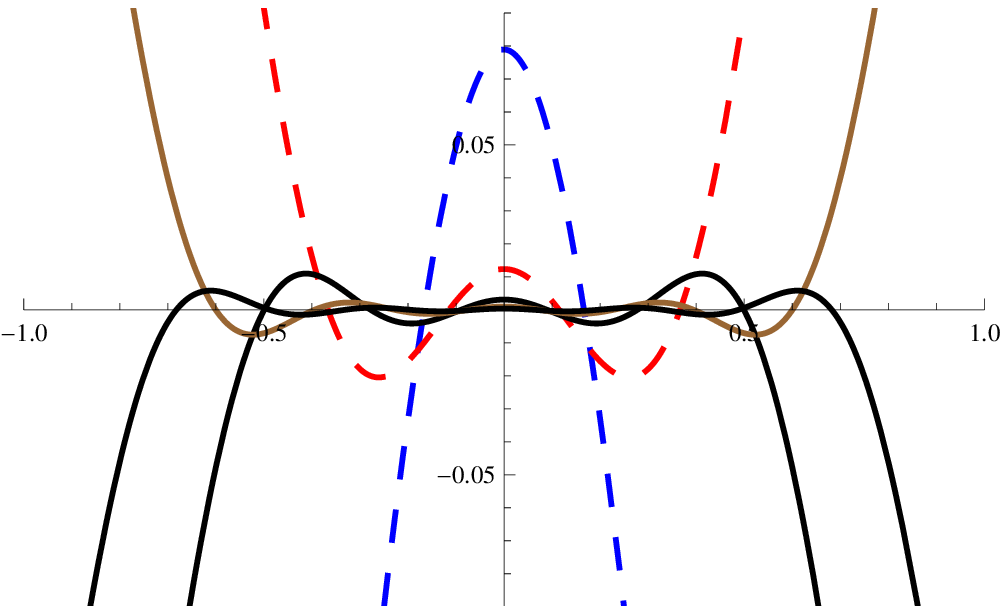}}
\caption{\label{neg30 normal}Symmetric solutions of the Heun equation ($\tilde{c}=-30$)
for $m = 0$ (black dash-doted), $m = 8.69355330$ (blue dashed), $m=13.16860126$ (red dashed),
$m=17.10643340$ (black solid), $m=20.80307154$ (brown solid),
$m=24.36269818$ (orange solid). Curves required masses with
thirty significant digits of which only the first are shown.
(\text{a}) Solutions eq.(\ref{hc_a4_s}) near the origin}
\end{figure}

\begin{figure}[ht]%
\centering
\includegraphics[height=4.5cm]{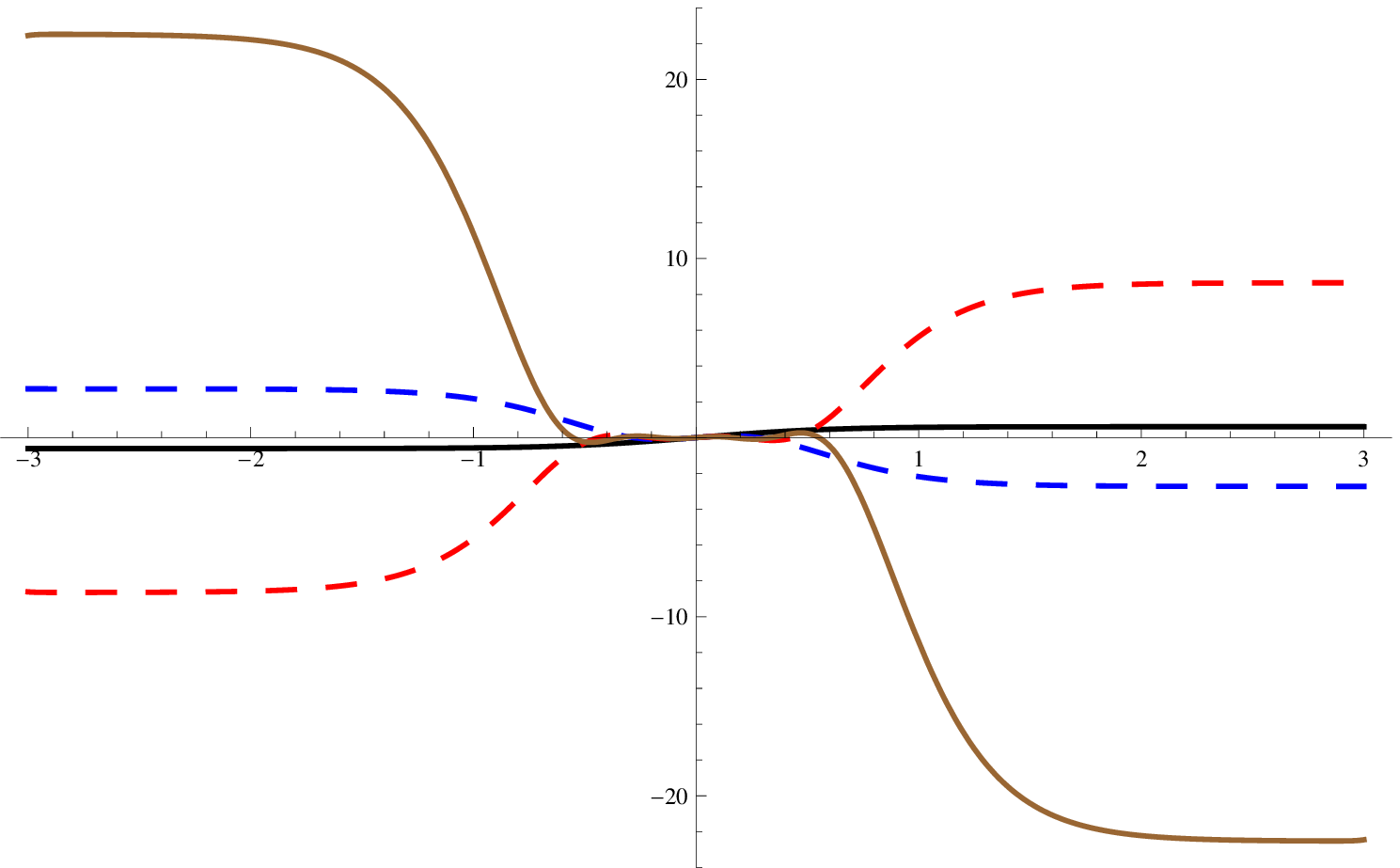}
\hspace{0.5cm}\subfigure[~]{
\includegraphics[width=4.0cm,height=3.0cm]{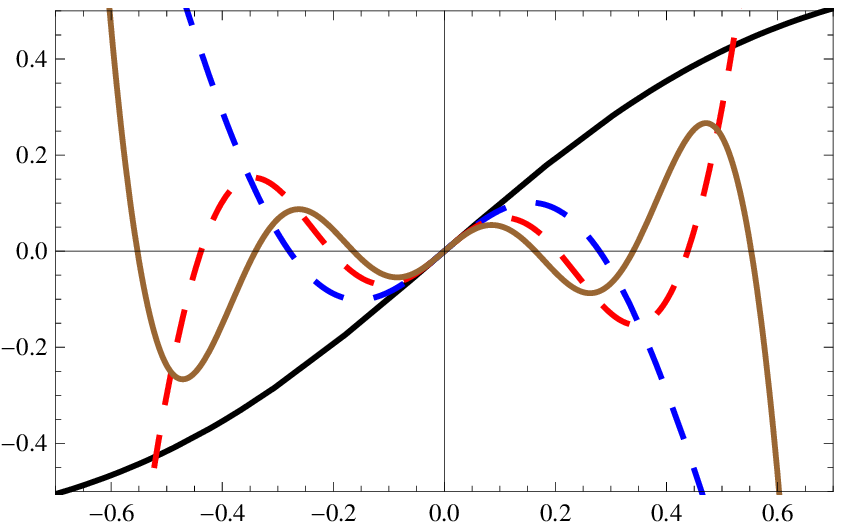}}
\caption{\label{neg30 anti}Antisymmetric solutions of the Heun equation ($\tilde{c}=-30$)
for $m = 5.90953031$ (black solid), $m = 11.03547247$ (blue dashed),
$m=15.17836781$ (red dashed), $m=18.97642242$ (brown solid).
(\text{a}) Solutions eq.(\ref{hc_a4_a}) near the origin.}
\end{figure}

A noteworthy point in the present approach is that now, depending on $\tilde{c}$,
the mass values $m^2$ can be quantized, as we saw in Sect. \ref{quantum analog}, or not,
as we will explain in what follows.

After a lengthy numerical exam, we found clear evidence that for $\tilde{c}\leq 0$
(namely $\lambda \geq \lambda_1\equiv 15/17 \lambda_0$) all the mass spectra are discrete.
For $\tilde{c}\in (0,4),\ \lambda_0< \lambda <\lambda_1$, on the other hand,
the corresponding spectra start with a zero mode and grow continuously.
This sharp contrast may be traced back to eq. (\ref{massive modes}) where the
second term of the differential equation
flips precisely with the sign of $\tilde{c}$.
Note that for any well-behaved solution $u(y)$,
the third term of eq. (\ref{massive modes})
can be disregarded at infinity. The remainder differential
equation can be easily solved showing that,
for $\tilde{c}>0$, solutions are always convergent to zero and
 for $\tilde{c}\leq0$ they diverge at the boundary. Guided by this result, we
performed a numerical survey in each region arriving at the conclusion above:
for $\tilde{c}>0$ there exist physical solutions for arbitrary $m$ while,
otherwise, only a discrete sequence of masses allow for finite solutions at the border.

It should be mentioned
that for small values of $\tilde{c}$ the solutions stabilize quickly. On the contrary,
for $\tilde{c}\lesssim -10$ the numerical calculation is more difficult and more
digits are needed in the mass precision to stabilize solutions at large values of $y$.
For example, for $\tilde{c}=-30$, which corresponds to $\lambda =0$, more than
thirty significant digits were necessary in the mass spectrum to find the solutions
as shown in Figs. \ref{neg30 normal} and \ref{neg30 anti}.
In Table \ref{table4} we listed the first values
of the mass up to the eighth decimal place.

\begin{figure}[ht]
\centering
\includegraphics[width=7.5cm,height=5.5cm]{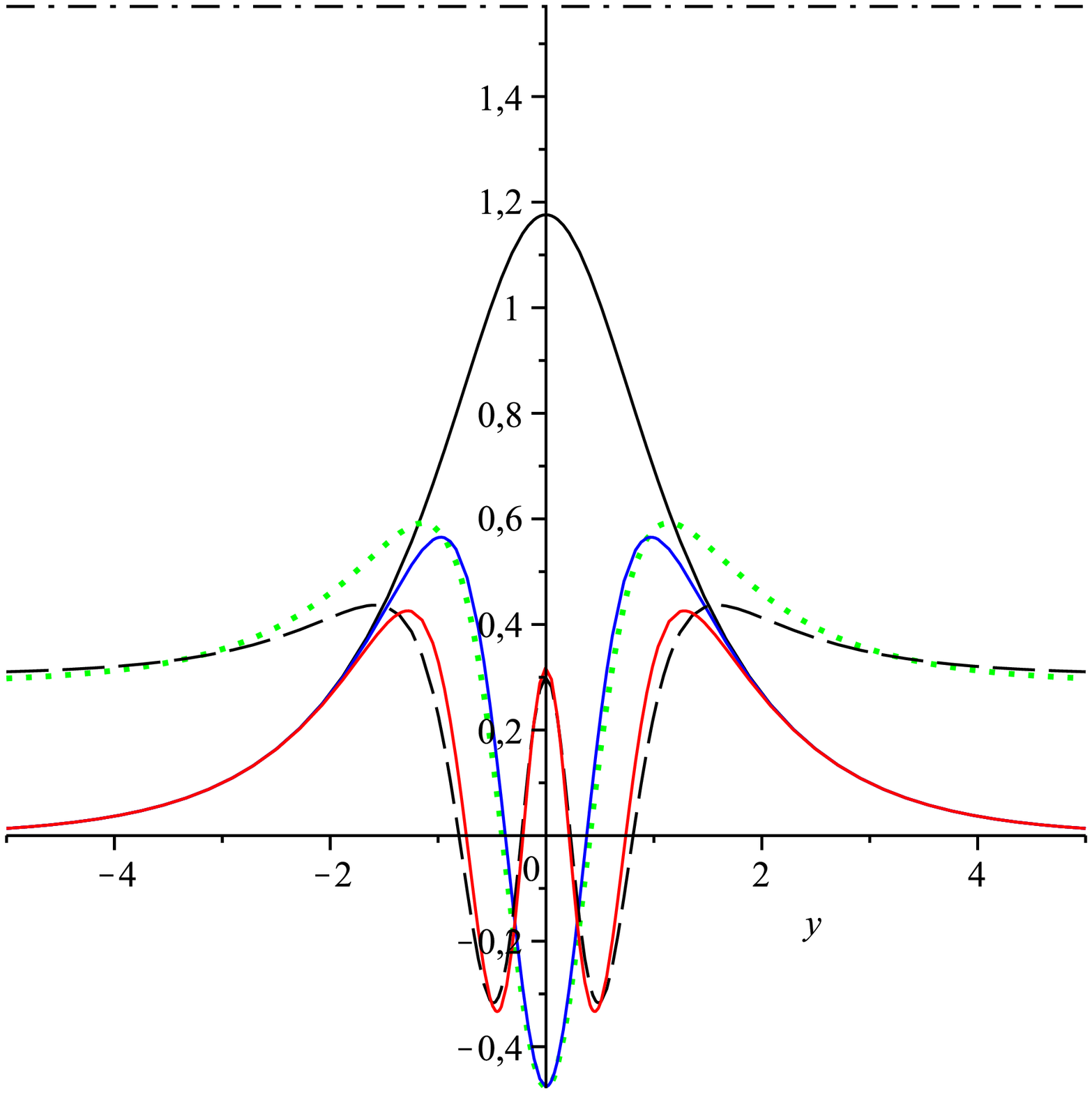}
\caption{\label{heun_sim_c_3_8} Symmetric solutions Eq. (\ref{hc_a4_s}) for $\tilde{c} =1$;
 $m=0$ (dash-dotted black line), $m=1.14718042$ (solid  black line), $m=4$ (dot green line), $m=4.30206964$ (solid blue line), $m=7$ (dashed black line), and $m=7.44879288$ (solid red line).}
% ASSIMETRICAS SEGUN SALEN INTERNAMENTE SIN CAMBIOS DE NORMALIZACION
\includegraphics[width=7.5cm,height=5.5cm]{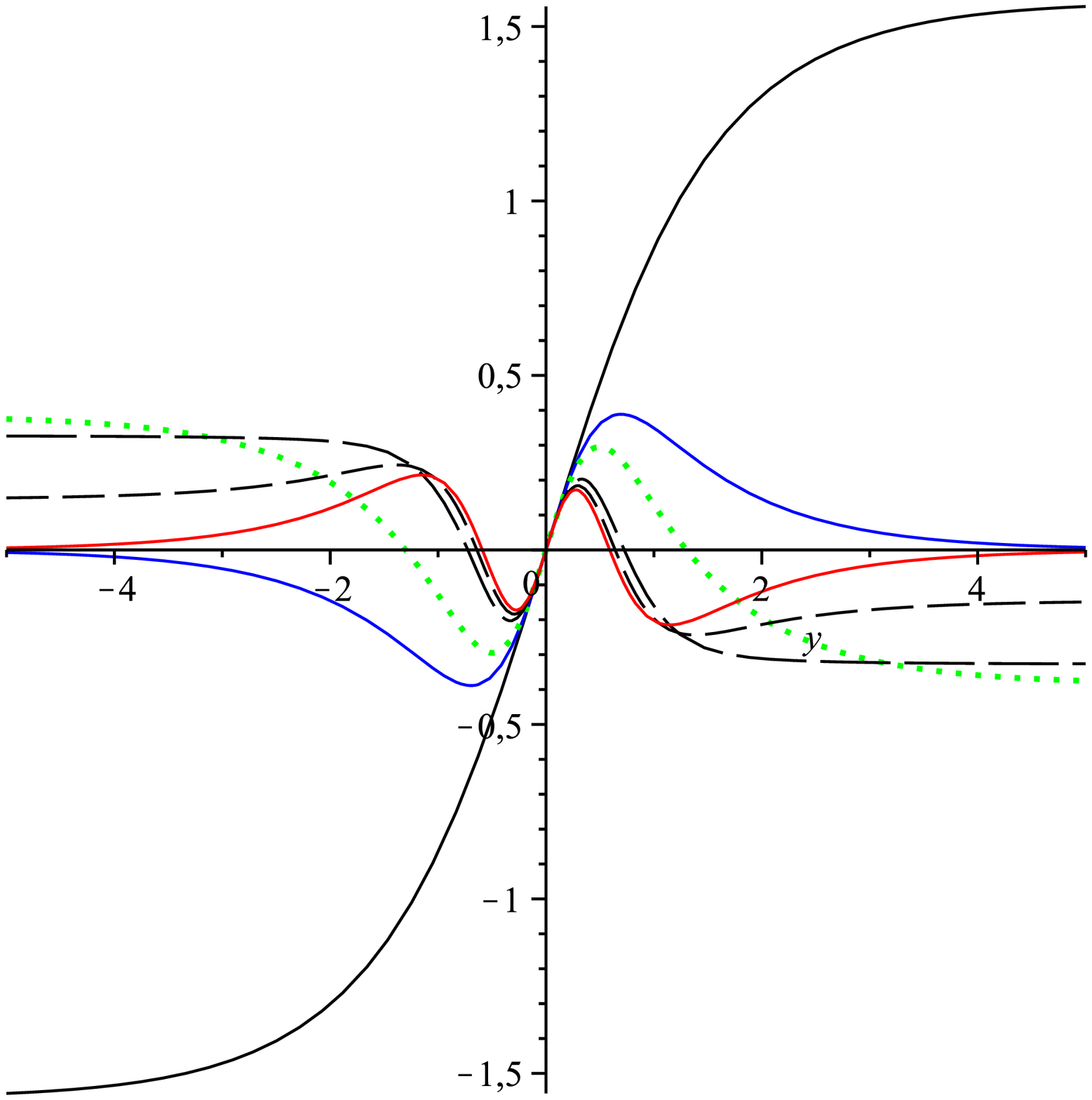}
\caption{\label{heun_asim_c_3_8} Antisymmetric solutions Eq. (\ref{hc_a4_a}) for $\tilde{c} =1$; $m=0$ (solid black line), $m=2.726324772$ (solid blue line), $m=3.5$ (dotted green line), $m=5$ (long-dashed black line), $m=5.5$ (dashed black line), and $m=5.875920066$ (solid red line).}
\end{figure}

\begin{figure}[ht]
\centering
\includegraphics[width=7.5cm,height=5.5cm]{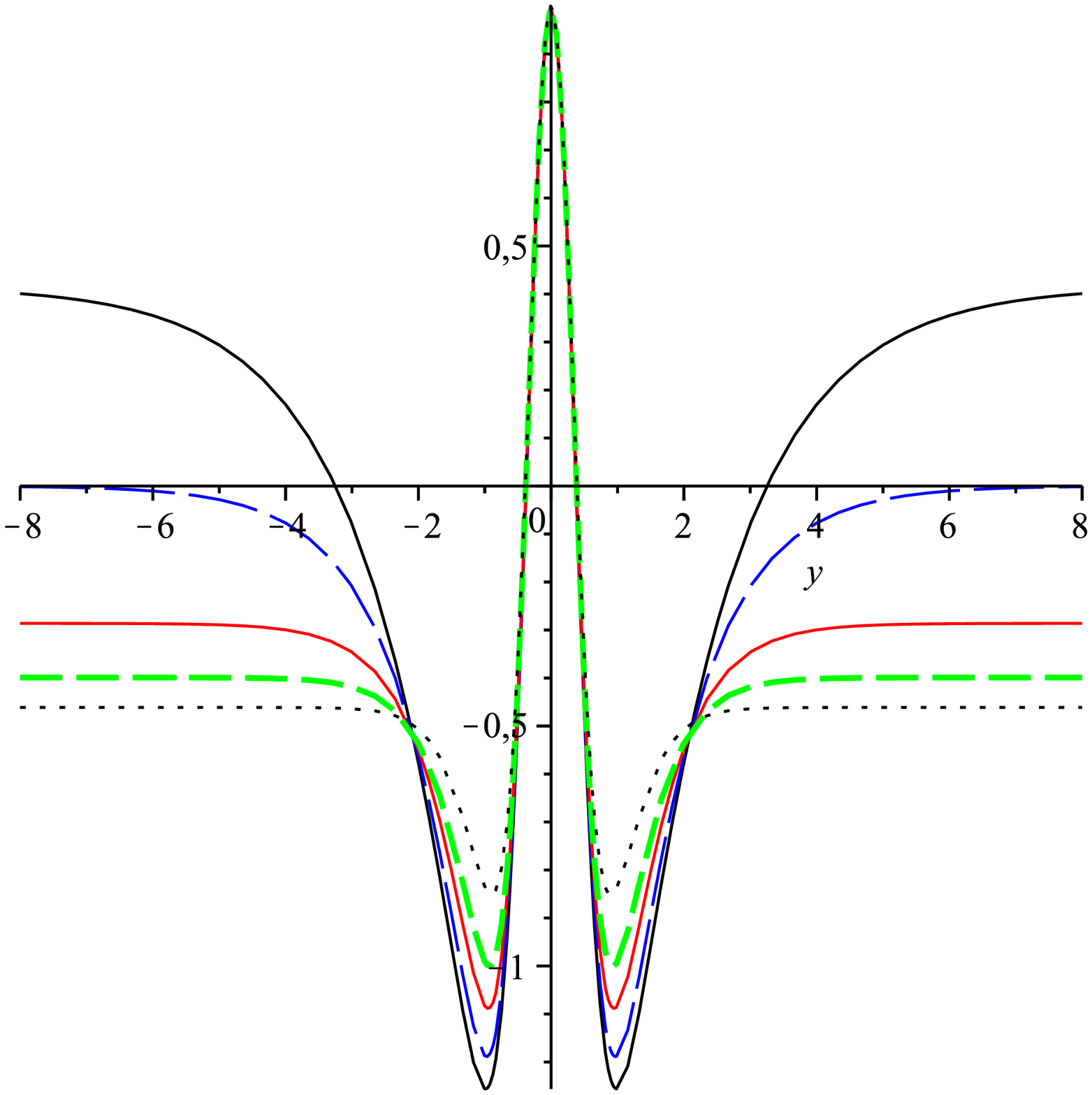}
\caption{\label{heun_sim_m_1} Symmetric Heun solutions,  Eq. (\ref{hc_a4_s}), for $m = 4.30206964$ and several values of $\tilde{c}$: $\tilde{c}=0.7$ (solid black line),  $\tilde{c}=1$ (long-dashed blue line), $\tilde{c}=1.5$ (solid red line), $\tilde{c}=2$ (dashed green line), and $\tilde{c}=3$ (dotted black line).}
%%%%%%%%%%%%%%%%%%%%%%%%%%%%%%%%%%%%%%%%%%%%%%%%%%%%%%%%%%%%%%%%%%%%%%%%%%%%%%%%%%%%%%%
% ASSIMETRICAS
%\includegraphics[width=7.5cm,height=5.5cm]{heun_asim_m_1.eps}
\includegraphics[width=7.5cm,height=5.5cm]{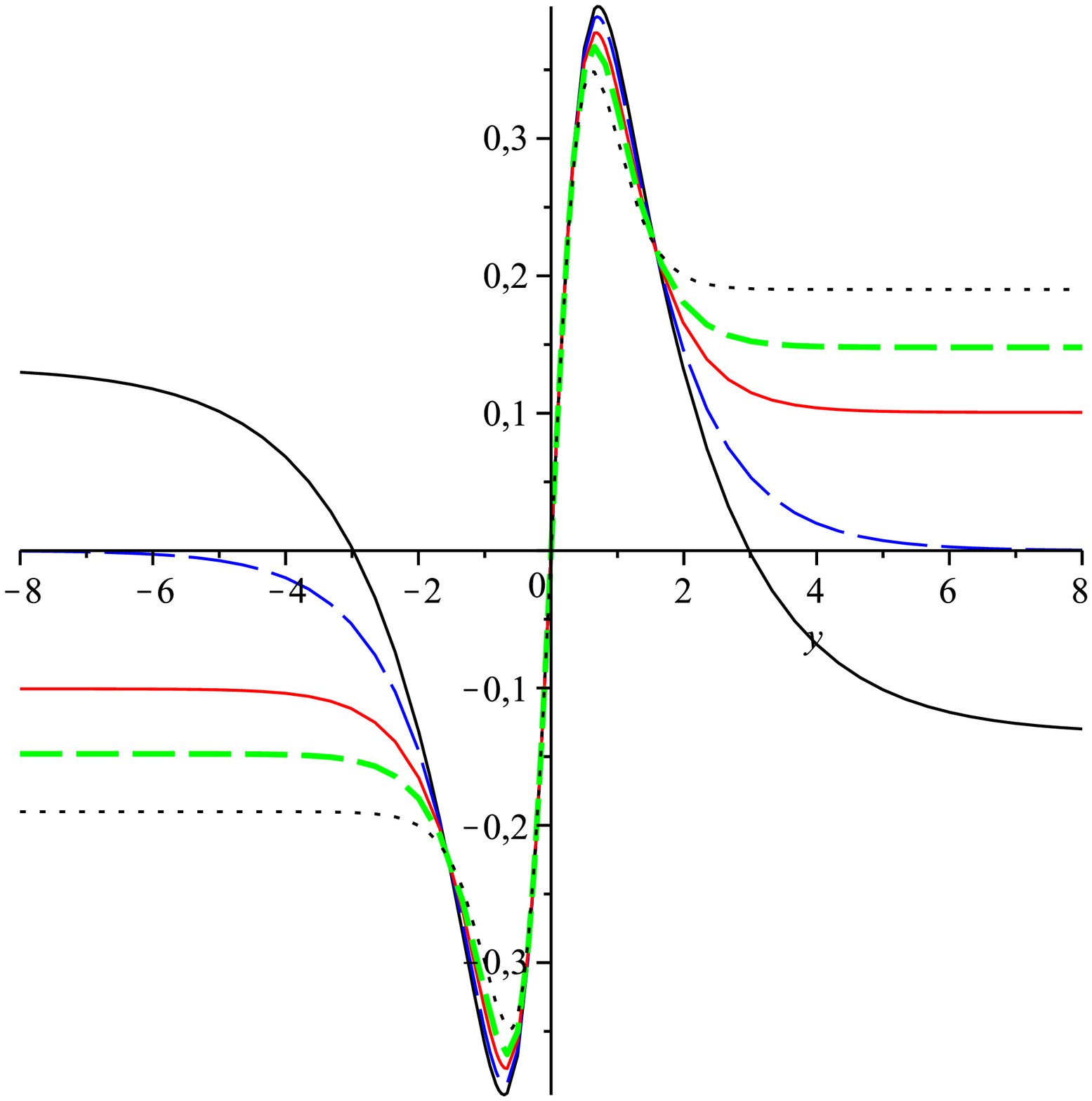}
\caption{\label{heun_asim_m_1} Antisymmetric Heun solutions, Eq. (\ref{hc_a4_a}), for $m = 2.72632477$ and several values of $\tilde{c}$: $\tilde{c}=0.7$ (solid black line),  $\tilde{c}=1$ (long-dashed blue line), $\tilde{c}=1.5$ (solid red line), $\tilde{c}=2$ (dashed green line), and $\tilde{c}=3$ (dotted black line).}
\end{figure}

As awaited, for the cases studied in the previous Section we find again the same results.
Note however that the zero-mode in the $\tilde{c}=1\ (c=3/8)$ case now
appears explicitly. This was expected since there exists an analytical $m=0$ solution to
Eq. (\ref{massive modes}), namely $u_0(y) = e_1 \arctan(e^y)+ e_2$, which must be present in a
full approach. Furthermore, for $\tilde{c}=1$
the Heun solution indicates that Table II would not only start from zero but
would also be continuously filled in as mentioned above.
In Figs. \ref{heun_sim_c_3_8} and \ref{heun_asim_c_3_8}
we can see finite analytic Heun solutions, given by eq. (\ref{hc_a4_s}) and eq. (\ref{hc_a4_a}),
for some arbitrary values of $m$ besides the quantum-mechanical analog ones. Another way to see it
is by means Fig. \ref{heun_sim_m_1} and Fig. \ref{heun_asim_m_1}
where $m$ has been fixed arbitrarily to one of the eigenvalues of $\tilde{c}=1$
and $\tilde{c}$ is then varied.
In the $\tilde{c}=-1\ (c=5/8)$ case the spectrum still obeys quantized values as given in Table I.

%%%%%%%%%%%%%%%%%%%%%%%%%%%%%%%%%%%%%%%%%%%%%%%%%%%%%%%%%%%%%%%%%%%%%%%%%%%%%%%%%%%%%%%
% FIGURAS HEUN c = 3/8, c~ = 1
%%%%%%%%%%%%%%%%%%%%%%%%%%%%%%%%%%%%%%%%%%%%%%%%%%%%%%%%%%%%%%%%%%%%%%%%%%%%%%%%%%%%%%%
%SIMETRICAS \textbf{DEBIDAMENTE RENORMALIZADAS} PARA COINCIDIR CON LAS MATHIEU

% FIGURAS HEUN m = 4.302, c~ = 0.7, 1, 1.5, 2, 3 -dentro de (0,4)
% ESTAS FIGURAS -NO- FUERON NORMALIZADAS COMO LAS ANTERIORES.
% SON TODAS HEUN CON UNA CONDICION INTERNA QUE IMPONE QUE SEAN TODAS 1 EN EL ORIGEN
%%%%%%%%%%%%%%%%%%%%%%%%%%%%%%%%%%%%%%%%%%%%%%%%%%%%%%%%%%%%%%%%%%%%%%%%%%%%%%%%%%%%%%%
%SIMéTRICAS

Thus, although Mathieu functions have been sufficient to characterize a part of the spectrum of the Schrodinger analog of our problem, we actually need to consider confluent Heun functions to cover all the cases.
In other words, even when we achieved fully analytical solutions
of the quantum analog differential equation, the spectra appeared just discrete not revealing
that some of them could be eventually continua.

The set of confluent Heun functions therefore provide
all the possible physical solutions of the actual problem in the 5D space.
This was not apparent from the Hamiltonian point of view which assumes the Sturm-Liouville operator $\mathcal{H}=\left[-\frac{d^2}{dz^2}+ \mathfrak{V}(z)\right]$ to represent the physical situation.

\section{Final remarks and conclusion\label{remarks}}

\begin{figure}
\includegraphics[width=7.5cm,height=5.5cm]{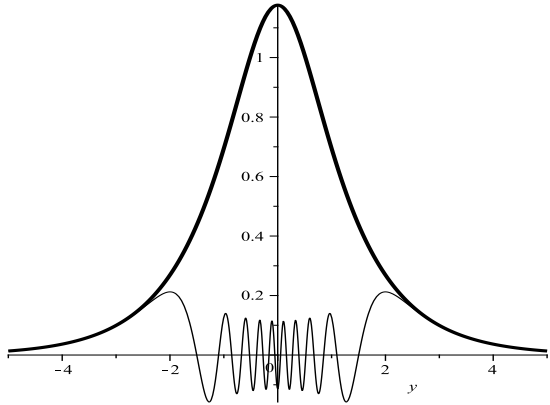}
%[modo high n]
\caption {\label{mathieu_c_3_8_m1-m29} First and tenth even KK eigenmodes
exhibit their relative weights. }
\end{figure}

\begin{figure}
\includegraphics[width=7.5cm,height=5.5cm]{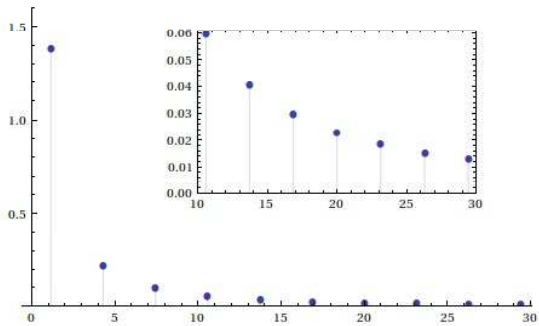}%{mathieurelativo.eps}
\caption {\label{mathieurelativo}Sequence of the first KK values of
$u^{2}_{\rm even}(0)$ for $c=3/8$ displaying the relative weights of
the KK modes on the brane.}
\end{figure}

In order to physically assess massive modes, one can evaluate
the variation of the effective gauge coupling as a function of  the Kaluza-Klein masses.
Actually, KK contributions can not be significant as compared with the Coulomb potential
because the coupling of massive modes to (fermion) matter on the brane develops a Yukawa 
type potential in the non-relativistic limit.
To show that this is a decreasing function of $m$ we should
evaluate the coefficients
that multiply the relevant sector of the four-dimensional action
\beq\sim\int dy\ {\rm e}^{\Sigma(y)-\lambda\Pi(y)/2}\left(u^2_{m=0}(y)+\sum_n\ u_{m_n}^2(y)\right)\int d^4x f_{\mu\nu}f^{\mu\nu}.\eeq
%Since these functions have divergent norms, such coefficients should be first renormalized.
%
However, in order to simplify this computation we can assume that the coupling with the brane takes place exactly on the 4D ordinary space-time, namely at $y=0$. It is precisely at this value of $y$ where the
relevant physical effects should be much stronger. For simplicity
let us consider the series of the quantum analog eigenvalues which serves as a discrete representative of the continuum. Thus, the effective 4D electrostatic potential would read
\beq V(r)\sim{q_1q_2}\left( \frac{c_0^2}{r}+\sum_{n} \frac{e^{-m_n\ r}}{r} u_{m_n}^2(0)\right)\eeq
where $q_1, q_2$ are two test charges separated a distance $r$
in ordinary 3D space and the Kaluza-Klein masses $m$ are numbered with $n$
in ascending order. See  Fig. \ref{heun_sim_c_3_8} where the first $u^{(1)}_{\rm even}(y)$
modes are fully displayed, and Fig. \ref{mathieu_c_3_8_m1-m29} where the first and the tenth modes are compared. See Fig. \ref{mathieurelativo} to appreciate the first 10 values at the origin. This, together with the negative exponential factor, essentially decouples the massive modes from the physics on the domain wall. Far from the membrane, all massive modes become constants like the zero-mode is, and as a consequence the 5D phenomenology results completely modified from ordinary 4D electromagnetism.  See e.g. Refs.\cite{RS, csaki al} for the study of this issue in the case of gravity.
\smallskip

\vskip 1cm
%{\bf Conclusion }

In this paper we have studied bulk and four-dimensional gauge propagation modes in a warped extra-dimensional space with a dilaton field.
We have set up a sine-Gordon thick membrane which bounces at the extra-coordinate origin. A five-dimensional metric was dynamically generated consistently
with the soliton brane and the dilaton background.
In such a framework we studied the solutions of a five-dimensional gauge field.

First, we have found the exact quantum-mechanical analog of our original five-dimensional stringy problem. We have shown that the corresponding  Schrodinger potential function is a quotient of simple second- and fourth-order polynomials that we could solve analytically.
We next obtained the exact quantum-mechanical analog eigenspectrum and used it as a guide
to analyze eventually the general solution.
A localized zero-mode corresponding to the ordinary photon was
guaranteed for a dilaton coupling constant above $\lambda_0$.
In general, we have found that the gauge-field dynamics are analytically given by confluent Heun functions which we have displayed for several representative cases.
Furthermore, in contrast to the quantum analog results, in the general approach the
mass of the gauge-field modes can be arbitrary for $\lambda \in(\lambda_0, \lambda_1)$. In any case,
we have shown that the Kaluza-Klein gauge spectrum is strongly attenuated on the brane
as compared to the zero-mode of the theory.
On the other hand, we observed that in the bulk, far from the brane, the amplitude of
an infinite tower of massive modes gets progressively relevant.
Interestingly, the quantum-mechanical discrete
mass eigenfunctions are completely decoupled in that region.

\end{document}